\documentclass[letterpaper,twocolumn,showpacs,aps,superscriptaddress,nofootinbib]{revtex4-2}

\usepackage[utf8]{inputenc}
\usepackage[english]{babel}

\usepackage{latexsym}
\usepackage{amssymb}
\usepackage{amsmath}

\usepackage{bm} 

\usepackage{epsfig}
\usepackage{graphicx}
\usepackage{float}

\usepackage{xcolor}

\usepackage{tensor} 

\usepackage[colorlinks=true,allcolors=blue]{hyperref}

\usepackage{comment}

\usepackage[normalem]{ulem}

\begin{document}


\title{Non-linear stability analysis of $\ell$-Proca stars}

\author{Claudio Lazarte}
\email{claudio.lazarte@uv.es}
\affiliation{Departamento de Astronom\'ia y Astrof\'isica, Universitat de Val\`encia,
Doctor Moliner 50, 46100, Burjassot (Val\`encia), Spain}

\author{Nicolas Sanchis-Gual}
\email{nicolas.sanchis@uv.es}
\affiliation{Departamento de Astronom\'ia y Astrof\'isica, Universitat de Val\`encia,
Doctor Moliner 50, 46100, Burjassot (Val\`encia), Spain}

\author{José A. Font}
\email{j.antonio.font@uv.es}
\affiliation{Departamento de Astronom\'ia y Astrof\'isica, Universitat de Val\`encia,
Dr. Moliner 50, 46100, Burjassot (Val\`encia), Spain}
\affiliation{Observatori Astronòmic, Universitat de València, C/ Catedrático José Beltrán 2, 46980, Paterna (València), Spain}

\author{Miguel Alcubierre}
\email{malcubi@nucleares.unam.mx}
\affiliation{Instituto de Ciencias Nucleares, Universidad Nacional
Aut\'onoma de M\'exico, Circuito Exterior Ciudad Universitaria, A.P. 70-543, M\'exico D.F. 04510, Mexico}




\begin{abstract} 

Vector boson stars, also known as Proca stars, exhibit remarkable dynamical robustness, making them strong candidates for potential astrophysical exotic compact objects.
In search of theoretically well-motivated Proca star models, we recently introduced the $\ell$-Proca star, a multi-field extension of the spherical Proca star, whose $(2\ell + 1)$ constitutive fields have the same time and radial dependence, and their angular structure is given by all the available spherical harmonics for a fixed angular momentum number $\ell$. In this work, we conduct a non-linear stability analysis of these stars by numerically solving the Einstein-(multi, complex) Proca system for the case of $\ell = 2$, which are formed by five constitutive independent, complex Proca fields with $m = 0, |1|$, and $|2|$. Our analysis is based on long-term, fully non-linear, 3-dimensional numerical-relativity simulations without imposing any symmetry. We find that ($\ell=2$)-Proca stars are unstable throughout their entire domain of existence. In particular, we highlight that less compact configurations dynamically lose their global spherical symmetry, developing a non-axisymmetric $\tilde{m}=4$ mode instability and a subsequent migration into a new kind of multi-field Proca star formed by fields with different angular momentum number, $\ell=1$ and $\ell=2$, that we identify as unstable multi-$\ell$ Proca stars.    
\end{abstract}




\maketitle


\section{Introduction}
\label{sec:intro}

Proca stars~\cite{BRITO2016291} are theoretical exotic compact objects composed of massive complex Abelian vector fields minimally coupled to Einstein's equations. Seen as macroscopic quantum condensates of massive spin-1 particles (vector bosons), they classify as bosonic stars. 
Compared to their scalar counterparts---the (scalar) boson stars first proposed in the late 1960s~\cite{Feinblum1968,Kaup68,Ruffini69} and extensively studied over the following decades---Proca stars have only been discovered relatively recently~\cite{BRITO2016291}. The limited attention given so far to vector fields may be attributed to the absence of massive Abelian vector fields in the Standard Model of particle physics, which instead incorporates a fundamental scalar field, the Higgs field. Nevertheless, while scalar fields have been motivated by theoretical frameworks such as the string axiverse~\cite{Arvanitaki:2009fg, Arvanitaki:2010sy}, recent extensions of the Standard Model have also introduced Proca fields as plausible candidates~\cite{freitas2021ultralight}.

Despite this, vector fields started to be seen as alternatives to the scalar case when it was found that complex Proca fields around Kerr black holes (BHs) also exhibit superradiant instabilities~\cite{Pani:2012_prl,Pani:2012bp,brito2020superradiance}. These findings motivated the construction of Kerr BHs with synchronized Proca hair solutions~\cite{Herdeiro:2016}.  
In addition, the first fully non-linear simulation of the end point of the superradiant instability was successfully performed with a complex Proca field~\cite{east2017superradiant}. Those simulations dynamically reached the solutions that correspond to fundamental states of the previously constructed hairy BHs~\cite{herdeiro2017dynamical, santos2020black}, thus, finding a dynamical formation mechanism of Kerr BHs with Proca hair. This remarkable result also connects with the theoretical existence of dynamically robust  rotating Proca stars, since in the limit of a vanishing BH horizon the solutions reduce to the rotating Proca stars discussed in~\cite{Sanchis_Gual_2019fae} (see also~\cite{BRITO2016291} for excited rotating Proca star solutions). Furthermore, hairy BHs are appealing from an astrophysical point of view because the timescale of the Proca field superradiant instability phase is much shorter than for the scalar case~\cite{Witek:2012tr}, by up to three to four orders of magnitude. This has been used to put tight constraints on the existence of such bosonic fields~\cite{cardoso2018constraining,palomba2019direct,siemonsen2020gravitational,tsukada2021modeling}.

Nowadays, the study of Proca stars is becoming increasingly relevant as these solutions offer simple models of astrophysical compact objects and serve as dark matter-motivated alternatives to black holes~\cite{cunha2023exotic,herdeiro2023black,herdeiro2024non}. Indeed, the high precision afforded by the increasing number of gravitational wave observations made by the LIGO-Virgo-KAGRA collaboration~\cite{abbott2019gwtc,abbott2021gwtc,abbott2023gwtc} has begun to open up the opportunity to test the BH paradigm. To explore this issue, research has been conducted on the emission of gravitational radiation from scalar~\cite{evstafyeva2024gravitational} and Proca star mergers~\cite{Sanchis_Gual_2019hop,Sanchis-Gual_2022, palloni2025eccentric}, enabling systematic searches of these mergers in the detected gravitational wave events \cite{Bustillo2021,bustillo2023svb,Bustillo2023inp}. Even if only as a proof of concept, such analyses have shown that gravitational-wave event GW190521~\cite{LIGO:gw190521} could be consistently interpreted as a head-on collision of Proca stars. This simplified setup (a head-on merger) motivates further investigations of compact objects beyond the Kerr BH paradigm. In light of these results and the ongoing advancements in observational capabilities, there is a growing urgency to deepen our understanding of exotic compact objects~\cite{bezares2025exotic}, prompting the search for well-motivated theoretical models of Proca stars.

Matter is \textit{quantum} at a fundamental level, so we can expect that some properties derived from the quantum formalism should also be observable in these hypothetical bosonic objects. In the scalar sector, the pioneering work of Ruffini and Bonazzola~\cite{Ruffini69} already considered a quantized scalar field where all bosonic particles, defined as excitations of such a field, were in the ground state forming a self-gravitating system once coupled with classical Einstein gravity. This work was recently extended in~\cite{Alcubierre:2023}, where, by preserving only the spherical symmetry and static nature of the classical spacetime, solutions composed of bosons in different excitation states were obtained. In addition, this work showed that, within the framework of semiclassical gravity in quantum field theory, a boson star composed of a single real quantum scalar field, with a large number of bosons occupying $N$ different excitation modes, can be equivalently described as a system of $N$ independent complex classical scalar fields.
This strongly indicates that, from a quantum perspective, it is more natural to investigate multi-field classical boson stars.

In this semiclassical formalism, several multi-field boson stars arise naturally; however, we are interested in a particular kind of solution that in the scalar sector has been proven to enjoy high dynamical stability. These are the $\ell$-boson stars~\cite{Alcubierre:2018}, spherically symmetric self-gravitating condensates where all the bosons have the same particle energy $E=\hbar \omega$ and angular momentum $L=\hbar\sqrt{\ell(\ell+1)}$ and are homogeneously populating all the available excitation modes for the fixed particle angular momentum number $\ell$, namely, modes with $m=-\ell,...,\ell$. Indeed, linear stability analysis of $\ell$-boson stars for different values of $\ell$~\cite{Alcubierre:2021} has shown that they are stable against spherical perturbations, while the dynamical time evolutions of these systems have revealed their robustness
under fully non-linear spherical perturbations for general $\ell$~\cite{Alcubierre:2019}, as well as non-linear non-spherical perturbations for the particular case of $\ell=1$~\cite{Jaramillo:2020}. Remarkably, numerical simulations in 3D performed in \cite{Sanchis-Gual:2021} demonstrated that $\ell$-boson stars with $\ell=0,1$ are the symmetry-enhanced points and unique stable configurations of a family of multi-field and multi-frequency boson stars comprising up to three constitutive fields. In alignment with this finding, recent reports have indicated that the ground states of Newtonian $\ell$-boson stars are only stable with $\ell=0,1$ under linear non-spherical perturbations~\cite{Chavez:2023}. These instabilities are probably present in the relativistic and non-linear scenario for $\ell\geq2$.

Given the astrophysical potential that single-field Proca stars have and following the quantum motivation to study multi-field bosonic stars, it is worth asking if there exists an analogous system to the $\ell$-boson stars in the vector scenario that also enjoys generic dynamical stability to be considered viable astrophysically. This paper is the second installment in a series addressing this inquiry. In our previous work~\cite{Lazarte:2024a}, we found $\ell$-Proca star solutions, as equilibrium configurations of spherically symmetric and static multi-field objects formed with $2\ell+1$ independent complex classical Proca fields. The goal of this paper is to conduct a thorough stability analysis of these solutions in non-linear regimes using 3D simulations without imposing any spatial symmetry.

This article is organized as follows. In Section~\ref{sec:system} we introduce the Einstein-(multi, complex) Proca system, its $\ell$-Proca star solutions together with a summary of the results obtained in~\cite{Lazarte:2024a} that will be useful for our following discussions, and its 3+1 decomposition expressed in a well-posed Cauchy problem.  Section~\ref{sec:numerical_setup} describes our initial data, our numerical setup for the simulations, and the details of the conducted convergence test for our evolutions. In Section~\ref{sec:stability_analysis} we present the results from our stability analysis, starting with the definition of the analysis quantities and following with the discussion of the identified different stages the $\ell$-Proca star evolutions undergo. We conclude in Section~\ref{sec:conclusions}. Throughout this paper we use the signature $(-,+,+,+)$ for the spacetime metric and Plank units such that $G = c = \hbar = 1$.


\section{The Einstein--(multi, complex) Proca system}
\label{sec:system}

\subsection{Field equations}

The Einstein-(multi, complex) Proca system that we use in this work is defined by an arbitrary number $N$ of complex, classical Proca fields $(X_m)_\alpha$, labeled with $m=1,...,N$, each of mass $m_V$ and minimally coupled to gravity. The action for this system is
\begin{equation}\label{Action:E-(multi)Proca}
S = \int d^4x \sqrt{-g}\ \left( \frac{R}{16\pi} + \mathcal{L}_{V} \right) \; ,
\end{equation}
where the Lagrangian density for the collection of Proca fields is
\begin{equation}\label{Lagrangian:multi-Proca}
    \mathcal{L}_{V} = - \frac{1}{16\pi} \sum^{N}_{m=1} \Big[ (\bar{W}_m)_{\mu\nu}(W_m)^{\mu\nu} + 2 \ m_V^2\ (\bar{X}_m)_\mu (X_m)^{\mu} \Big]\; .  
\end{equation}  
Here, $W_m$ denotes the Proca field strength of the $m$th Proca field, given by $(W_m)_{\mu\nu} = \nabla_\mu (X_m)_\nu - \nabla_\nu (X_m)_\mu$, and the complex conjugates are denoted by an overbar. Notice that this Lagrangian density is endowed with global internal symmetries.  In fact, in addition to its invariance under $U(1)$ transformations resulting from the complex character of each Proca field,  since these $N$ fields have the same mass  and they do not directly interact between themselves, Eq.~\eqref{Lagrangian:multi-Proca} is also invariant under 
$U(N)$ transformations in the internal field space. By applying the variational principle to Eq.~\eqref{Lagrangian:multi-Proca}, the equations of motion for the gravitational field are the Einstein equations $R_{\mu\nu}-\frac{1}{2} \: R\ g_{\mu\nu} = 8 \pi T_{\mu\nu}$, where the stress-energy tensor is defined as
\begin{equation}\label{def:stress-energy}
    T_{\mu\nu}:= \frac{-2}{\sqrt{-g}}\frac{\delta S_{\rm matter}}{\delta g^{\mu\nu}}\;,
\end{equation}
with $S_{\rm matter}=\int d^4x\sqrt{-g}\;\mathcal{L}_{V}$, which leads to
\begin{align}\label{exp:tensor:stress-energy}
T_{\mu\nu} & = \frac{1}{4\pi} \sum^N_{m=1} \left\{ -(W_m)_{\lambda(\mu}(\bar{W}_m)_{\nu)}^\lambda - \frac{g_{\mu\nu}}{4} (W_m)_{\alpha\beta}(\bar{W}_m)^{\alpha\beta} \right. \nonumber \\ 
&  + \left. m_V^2 \left[ (X_m)_{(\mu}(\bar{X}_m)_{\nu)} - \frac{g_{\mu\nu}}{2} \: (X_m)_\lambda (\bar{X}_m)^\lambda \right] \right\} \; .
\end{align}
The corresponding equations of motion for the matter fields are the Proca equations,
\begin{equation}\label{eq:field:Proca}
\nabla_\mu (W_m)^{\mu\nu} - m_V^2 (X_m)^\nu = 0 \; .
\end{equation}
Although each massive vector field $(X_m)_\mu$ does not possess gauge freedom, it can be shown from Eq.~\eqref{eq:field:Proca} that it must satisfy the Lorenz condition
\begin{equation}\label{eq:Lorenz}
\nabla_\nu (X_m)^\nu = 0 \; .
\end{equation}

\subsection{$\ell$-Proca stars}\label{subsec:ell-Proca_stars}
In the spherically symmetric sector, the Einstein-(multi) Proca system~\eqref{Action:E-(multi)Proca} admits static spacetime solutions when $N=1$. These are the standard Proca stars~\cite{BRITO2016291} made of one complex classical field. The spacetime staticity is induced by the internal symmetry group $U(1)$ of the matter field, since it allows one to express its time dependence in terms of complex phases given by $e^{-i\omega t}$, where $\omega$ is a real frequency parameter. Thus, in the Lagrangian density~\eqref{Lagrangian:multi-Proca}, the time variable is canceled by products of the fields with their complex conjugates. These products are invariant under time translations, leading to a time-independent stress-energy tensor that sources a static spacetime. It is important to mention that to achieve the spherical symmetry of these single-field Proca stars, we have to restrict the spatial dependence of the matter field to the radial coordinate. An angular dependence would lead to axially symmetric equilibrium configuration, as for instance, the static dipolar or quadrupolar Proca stars with $\theta$ dependence~\cite{herdeiro2024non} or the spinning Proca stars with $(\theta,\varphi)$ dependence~\cite{BRITO2016291,herdeiro2019asymptotically}. 

In the quest to extend spherical Proca stars to $N$-field configurations in the most general way possible, we can allow an angular dependence in the Proca fields, but hiding it in the Lagrangian density of the system in a similar way as we explained above for the time dependence. This is equivalent to requiring the products in~\eqref{Lagrangian:multi-Proca}, such as $\sum^{N}_{m=1}(\bar{X}_m)_\mu (X_m)^{\mu}$, to be invariant under spatial rotations. It is well known, from the group representation theory, that the spherical harmonics $Y^{\ell m}(\theta,\varphi)$ of degree $\ell$ provide the vector basis components of a $(2\ell+1)$-dimensional vector space that transforms by irreducible representations of the group of spatial rotations $SO(3)$. Therefore, under rotations of our spatial coordinates, the $Y^{\ell m}$ transform in such a way that products involving all the $2\ell+1$ spherical harmonics and their contracted covariant derivatives $\hat{\nabla}$ associated with the unit two-sphere metric $\hat{g}=\rm{diag}(1,\sin^2{\theta})$, such as $\sum^{\ell}_{m=-\ell}\bar{Y}^{\ell m}Y^{\ell m}$, $\sum^{\ell}_{m=-\ell}\hat{\nabla}^A\bar{Y}^{\ell m}\hat{\nabla}_A Y^{\ell m}$, or $\sum^{\ell}_{m=-\ell}\hat{\nabla}^B\hat{\nabla}^A\bar{Y}^{\ell m}\hat{\nabla}_B\hat{\nabla}_A Y^{\ell m}$, are invariant. To achieve spherical symmetry by constructing the above products in~\eqref{Lagrangian:multi-Proca}, the system must be first composed of $N=2\ell+1$ Proca fields and then, each of these fields with their angular dependence given by each $Y^{\ell m}$ must have the same radial amplitude to factorize it out of those products. Following these requirements, we proposed in~\cite{Lazarte:2024a} the ansatz of each $m$th Proca field in~\eqref{Lagrangian:multi-Proca}, expressed as a potential 1-form, as follows:
\begin{align}
X_m & =  \left[\; -F_\ell(r) Y^{\ell m} dt + i a_\ell(r) Y^{\ell m} dr \right. \nonumber \\ 
& \ \ \ \ \ +\left. i b_\ell(r) \left( \partial_\theta Y^{\ell m} d \theta 
+ \partial_\varphi Y^{\ell m} d \varphi \right)\;  \right] e^{-i \omega t} \; ,
\label{ansatz:potential:1-form}
\end{align}
where each field is from now on labeled by the parameter $m = -\ell,...,\ell\;$, and the real-valued time frequency $\omega$ and radial amplitudes $\{F_\ell,a_\ell,b_\ell\}$ are the same for all values of $m$. Notice that this expression reduces to the standard spherical Proca star 1-form~\cite{BRITO2016291} in the case of $\ell=0$. Furthermore, for $\ell\neq0$ values, since the constitutive fields with $m\neq0$ have the azimuthal dependence given by $e^{im\varphi}$, these fields are spinning Proca fields with a non-zero angular momentum, that by their own would be sourcing a stationary axially symmetric spacetime. However, the gravitational coupling of the $2\ell$ spinning fields cancels the total angular momentum, ensuring that the sourced spacetime is static. Then, the spherical symmetry is attained adding the coupling with the $(m=0)$-field, as we foresee from the beginning of the ansatz construction. The idea of extending the internal symmetry group of a bosonic star from $U(1)$ to $U(2\ell+1)$ in order to source more general spherically symmetric stars was inspired by a similar idea used in the context of gravitational collapse by Olabarrieta {\em et al.}~\cite{Olabarrieta:2007di}. The first case of applicability was in the context of scalar boson stars~\cite{Alcubierre:2018}, whose resulting extensions were called ``$\ell$-boson stars". Our ansatz~\eqref{ansatz:potential:1-form} stands for the application of this idea to the vector case. For that reason, we refer to the resulting bosonic stars as $\ell$-Proca stars. 

Despite the fact that the derivation of~\eqref{ansatz:potential:1-form} based on symmetries of the covariant form of Proca fields is possible, as we have shown, we approached the problem more pragmatically in~\cite{Lazarte:2024a}. We applied a 3+1 decomposition to the matter fields (introduced in the next subsection of this paper) to analyze the effects of different ansätze on the components of the stress-energy tensor. This approach allowed us to achieve the desired spherical symmetry of the system while simultaneously satisfying the ``Gauss constraint" of Proca theory (see below for its definition). Precisely, this is the purpose of the spatial dependence of the angular vector components chosen in our ansatz in Eq.~\eqref{ansatz:potential:1-form}, which introduces an additional radial degree of freedom $b(r)$ and angular dependence in terms of the angular derivatives of spherical harmonics.\footnote{We note that our ansatz is consistent with the expansion of an arbitrary vector field in the vector spherical harmonics basis. This allows a correct treatment on the spatial variable separability of Poisson equation, as shown in~\cite{Barrera:1985}, which in our case turns out to be crucial to satisfy the Gauss constraint.}  

As we can notice, the existence of $\ell$-Proca stars as solutions of the static and spherically symmetric Einstein-(multi, complex) Proca system is determined by the specification of the time frequency $\omega$, the radial profiles $\{F_\ell,a_\ell,b_\ell\}$, and the metric functions. These quantities are obtained using our ansatz to solve the  Proca and Einstein equations assuming a spherically symmetric spacetime given, in isotropic coordinates, by
\begin{equation}\label{metric:spherically-symmetryc}
    ds^2 = -\alpha^2(r)dt^2 + \psi^4(r)(dr^2+r^2d\Omega^2)\; ,
\end{equation}
where $\alpha$ is the lapse and $\psi$ is the conformal factor of the 3-metric. Taking advantage of the harmonic time dependency of the Proca fields, these equations of motion reduce to a system of five radial ordinary differential equations. Imposing appropriate asymptotic conditions  $F_\ell(\infty)=a_\ell( \infty)=b_\ell(\infty)=0
$ and $\alpha(\infty)=\psi(\infty)=1$, they form a non-linear eigenvalue problem for $\omega$. For an arbitrary and fixed $\ell$, this system reads as follows:
\begin{align}
0 &=  F''_\ell + \frac{2 F'_\ell}{r} - \frac{\ell(\ell+1)}{r^2} \: F_\ell \nonumber \\ 
& + \frac{2\psi'}{\psi} \: F'_\ell - \frac{\alpha'}{\alpha} \left( F'_\ell+2\omega a_\ell \right) - \psi^4 \left( m_V^2 - \frac{\omega^2}{\alpha^2} \right) F_\ell \; , \label{eq:initial_data:second_order:F}  \\[5.0pt]
 0 &=  \ a''_\ell + \frac{2a'_\ell}{r} - \frac{2a_\ell}{r^2} - \frac{\ell(\ell+1)}{r^2}a_\ell + \frac{2\ell(\ell+1)}{r^3}b_\ell  \nonumber \\ 
 & -\frac{2 a_\ell}{r}\left(\frac{\alpha'}{\alpha} + \frac{6\psi'}{\psi}\right) + \frac{4\ell(\ell+1)}{r^2}\frac{\psi'}{\psi}b_\ell +\frac{2\psi^4}{\alpha^2}\frac{\alpha'}{\alpha}\omega F_\ell  \nonumber \\
& - a_\ell \left[ 10\left(\frac{\psi'}{\psi}\right)^2 + \frac{6\alpha'\psi'}{\alpha\psi} + \left(\frac{\alpha'}{\alpha}\right)^2 \right] + \left(\frac{\alpha'}{\alpha} - \frac{2\psi'}{\psi}\right)a'_\ell \nonumber \\
& + 4\pi\psi^4 a_\ell S - \psi^4\left(m_V^2 - \frac{\omega^2}{\alpha^2}\right)a_\ell \; , \label{eq:initial_data:second_order:a}\\[5.0pt]
0 &= b''_\ell - \frac{\ell(\ell+1)}{r^2} \: b_\ell + \frac{2 a_\ell}{r} \nonumber \\ 
& + \left( \frac{\alpha'}{\alpha}-\frac{2\psi'}{\psi} \right) b'_\ell + \frac{4\psi'}{\psi} a_\ell
- \psi^4 \left( m_V^2 - \frac{\omega^2}{\alpha^2} \right) b_\ell \; ,\label{eq:initial_data:second_order:b} \\[3.5pt]
   0 &= \psi'' + \frac{2\psi'}{r} + 2 \pi \psi^5 \rho \; , \label{eq:initial_data:second_order:psi}\\
0 &= \alpha'' + \frac{2\alpha'}{r} +\frac{2\alpha'\psi'}{\psi}
- 4 \pi \alpha \psi^4 \left( S + \rho \right) \; , \label{eq:initial_data:second_order:lapse}
\end{align}
where $\rho$ and $S$ are, respectively, the energy density and the trace of the spatial stress tensor, matter quantities obtained from the orthogonal decomposition of the stress-energy tensor as we indicate in the next section.

To fully solve the second-order differential system~\eqref{eq:initial_data:second_order:F}-\eqref{eq:initial_data:second_order:lapse}, an additional set of five boundary conditions is required. In order to obtain non-singular configurations one needs regularity conditions that can be obtained analytically solving the above system in a small vicinity around the center.  Thus, for the metric functions, we have $\alpha = \alpha_0 + \mathcal{O}(r^2)$ and $\psi = \psi_0 + \mathcal{O}(r^2)$, while for the radial degrees of freedom of matter, it was found in \cite{Lazarte:2024a} that the local solutions near the origin have to be classified for trivial and non-trivial values of $\ell$. Then, for $\ell=0$, we need only one independent parameter $c_1$, and we have $ F_{\ell=0}=c_1 + \mathcal{O}(r^2)$ and $ a_{\ell=0}=c_1\left(\frac{\psi^4_0 \omega}{2\alpha^2_0}\right)r+\mathcal{O}(r^3)$, while, for $\ell\neq0$, the number of independent parameters to build regular general solutions are given by two parameters $c_1$ and $c_2$,
\begin{align}
    F_{\ell}&=c_1r^\ell + \mathcal{O}(r^{\ell+2})\;, \label{expr:local:F} \\ a_{\ell}&=\ell c_2r^{\ell-1} + \mathcal{O}(r^{\ell+1})\;, \label{expr:local:a}  \\ b_{\ell}&=c_2 r^\ell + \mathcal{O}(r^{\ell+2})\;. \label{expr:local:b} 
\end{align}
Furthermore, we note that the eigenvalue $\omega$ was also taken as an unknown variable by adding an extra equation that fixes the central value of one the radial functions of the system. For the equilibrium configurations studied in this work, we used the lapse function $\alpha$. 

The above system is solved using a multi-domain collocation spectral method~\cite{grandclement2009:spectralreview,Grandclement2014:rotatingBSs,grandclement2010:kadath,Alcubierre:2022} based on spectral projections onto a finite basis of Chebyshev polynomials with their basis parity restricted to impose the above regularity conditions. We also implemented the Galerkin technique~\cite{grandclement2010:kadath} to obtain the correct $r^\ell$ dependency of the radial functions at small radii, for $\ell=2,3$. (For details see~\cite{Lazarte:2024a}.) By applying this procedure, we obtained the radial profiles and the time frequency of the ground-state equilibrium configurations for $\ell$-Proca stars with $\ell=0,1,2,3$. 

In this work we are interested to evolve $\ell$-Proca stars for non-trivial values of $\ell$, as ($\ell=0$)-Proca stars were already evolved in~\cite{herdeiro2024non} and were found to be unstable against non-spherical perturbations.
The solutions of $\ell$-Proca stars with $\ell=1$ present a few difficulties. From the analytical local behavior of the matter radial functions~\eqref{expr:local:F}--\eqref{expr:local:b} around the origin, one can observe that for the case $\ell=1$, the function $a_\ell$ takes a constant value at the origin, namely $a_{\ell=1}=c_2$. This was also proven in~\cite{Lazarte:2024a} by observing its radial profile obtained numerically. As we can see from Eq.~\eqref{ansatz:potential:1-form}, $a_ \ell$ corresponds to the radial component of the 3-vector potential. Having a non-zero value at the origin means that the vector field has a discontinuity in this point (this is easily understood once the angular dependency is included to that component and we computed the value of the 3-vector potential field in the origin with different angles).  This lack of regularity at the origin led us to rule out the $(\ell=1)$-Proca star as a physical solution and start the stability analysis with the $\ell=2$ case. 

\subsection{3+1 decomposition}\label{sec:3+1decomposition}

To perform time evolutions we employ the 3+1 decomposition of the spacetime and the matter fields. The former is implemented as presented in~\cite{Alcubierre08a}, by foliating the spacetime into space-like hypersurfaces $\Sigma_t$ with 3-metric $\gamma_{ij}$, and the latter by splitting the Proca fields into scalar $\Phi_m$ and 3-vector $(a_m)_i$ potentials
\begin{equation}
\label{3+1:def:potentials}
\Phi_m := -n^\mu (X_m)_\mu \; , \quad
 (a_m)_i := \tensor{\gamma}{^\mu_i} (X_m)_\mu \; ,
\end{equation}
and defining 3-dimensional ``electric'' and ``magnetic'' fields as follows~\cite{Alcubierre:2009ij,Zilhão:2015}:
\begin{equation}
    \label{3+1:def:electricomagnetic_fields}
(\mathcal{E}_m)_i := -n^\mu\tensor{\gamma}{^\nu_i}(W_m)_{\mu\nu} \; , \quad
 (\mathcal{B}_m)_i := -n^\mu\tensor{\gamma}{^\nu_i}(W^*_m)_{\mu\nu} \; .
\end{equation}
In these equations $n^\mu$ is the unit time-like normal vector to the spatial hypersurfaces $\Sigma_t$, $\gamma^\mu_\nu := \delta^\mu_\nu + n^\mu n_\nu$ is the projection operator onto $\Sigma_t$, and $(W^*_m)^{\mu\nu} := - E^{\mu\nu\alpha\beta}(W_m)_{\alpha\beta}/2$  denotes the Hodge dual of the field strength tensor.\footnote{This definition is using the convention $E^{0123} = -1/\sqrt{-g}$ and $E_{0123} = \sqrt{-g}$.}

In the 3+1 formalism the stress-energy tensor, given by~\eqref{exp:tensor:stress-energy}, is also decomposed by the following projections: the energy density $\rho:=n^\mu n^\nu T_{\mu\nu}$, the momentum density $j_i := - \tensor{\gamma}{^\mu_i}n^\nu T_{\mu\nu}$, and the spatial stress tensor $S_{ij}:= \tensor{\gamma}{^\mu_i}\tensor{\gamma}{^\nu_j}T_{\mu\nu}$. Their form in terms of~\eqref{3+1:def:potentials}--\eqref{3+1:def:electricomagnetic_fields} is
\begin{align}
\rho = \sum^N_{m=1} \frac{1}{8\pi}&\left\{\, (\mathcal{E}_m)_i (\bar{\mathcal{E}}_m)^i + (\mathcal{B}_m)_i (\bar{\mathcal{B}}_m)^i \right. \nonumber \\ 
& \left. +\,  m_V^2\; [\Phi_m \bar{\Phi}_m + (a_m)_i(\bar{a}_m)^i ]\ \right\} \label{exp:3+1:Proca_energy_density}\; , \\
j^i = \sum^N_{m=1} \frac{1}{8\pi} &\left\{\, \tensor{E}{^i_{jk}}(\bar{\mathcal{E}}_m)^j (\mathcal{B}_m)^k + m_V^2\: (a_m)^i \: \bar{\Phi}_m \right. \nonumber \\ 
& \left. +\,  c.c.\right\} \label{exp:3+1:Proca_momentum_density} \; , \\
S_{ij} = \sum^N_{m=1}\frac{1}{8\pi}&\left\{\ \gamma_{ij}[\ (\mathcal{E}_m)_k (\bar{\mathcal{E}}_m)^k + (\mathcal{B}_m)_k (\bar{\mathcal{B}}_m)^k \ ] \right. \nonumber \\
&  - [ \ (\mathcal{B}_m)_i (\bar{\mathcal{B}}_m)_j + (\mathcal{E}_m)_i (\bar{\mathcal{E}}_m)_j + c.c. \ ] \nonumber \\[5pt]  
&  + \left. m_V^2\: [\ ( (a_m)_i(\bar{a}_m)_j + c.c. )\right. \nonumber \\[3pt] 
& \left.  - \, \gamma_{ij}( (a_m)_k(\bar{a}_m)^k  - \Phi_m \bar{\Phi}_m)\ ] \ \right\}  \label{exp:3+1:Proca_stress_tensor} \; .
\end{align}
Regarding the spacetime variables, we can obtain their evolution and constraint equations from projections of the Einstein equations. 
We employ the strongly hyperbolic Baumgarte-Shapiro-Shibata-Nakamura (BSSN) formulation~\cite{Shibata95, Baumgarte:1998te} with  the ``1+log" slicing for the lapse~\cite{Bona95} and the ``$\Gamma$-driver" gauge condition for the shift~\cite{alcubierre2003gauge}. This guarantees the well-posedness of the system~\cite{beyer2004well} and allows for robust and stable numerical simulations~\cite{Alcubierre:2018}. In this formulation, the 3-metric is conformally decomposed and the 4-dimensional line element reads 
\begin{equation}\label{4-metric}
    ds^2 = -(\alpha^2-\beta^i\beta_i)dt^2 + 2\beta_i dt dx^i + e^{4\phi}\tilde{\gamma}_{ij}dx^i dx^j \; , 
\end{equation}
where $\alpha$ is the lapse, $\beta^i$ is the shift vector, $\tilde{\gamma}_{ij}$ is the conformal 3-metric, and $\phi$ is the logarithmic conformal factor of the 3-metric that fulfills $\phi=\frac{1}{12}\ln\gamma$, with $\gamma$ the determinant of the 3-metric. In particular, and only for the spacetime evolution, we used the $W$ method~\cite{Marronetti2008high}, which instead of evolving $\phi$, evolves $W:=e^{-2\phi}$. Thus, in addition to this variable, our BSSN system evolves the conformal 3-metric $\tilde{\gamma}_{ij}$, the trace $K$ of the extrinsic curvature, the conformal trace-free extrinsic curvature $\tilde{A}_{ij}:=W^2(K_{ij}-\gamma_{ij}K/3)$, and the conformal connection functions $\tilde{\Gamma}^i:=\tilde{\gamma}^{jk}\tilde{\Gamma}^i_{jk}$,  using the following equations:
\begin{align}
    \frac{d}{dt} W  &= \frac{1}{3} \alpha W K\; , \label{eq:bssn:conformal_factor_W}\\[2.5pt]
    \frac{d}{dt} \tilde{\gamma}_{ij} & = -2\alpha\tilde{A}_{ij}\; , \label{eq:bbsn:conformal_metric} \\[2.5pt]
    \frac{d}{dt}  K &= - D^i D_i \alpha + \alpha \left(\tilde{A}_{ij}\tilde{A}^{ij} + \frac{1}{3}K^2 \right) \nonumber \\
    & + 4\pi\alpha(\rho+S)\; ,\label{eq:bssn:trace_extrinsic_curv} \\[2.5pt]
    \frac{d}{dt}\tilde{A}_{ij} &= W^2(-D_iD_j\alpha + \alpha R_{ij})^{\rm TF} + \alpha(K\tilde{A}_{ij} - 2\tilde{A}_{ik}\tensor{\tilde{A}}{^k_j}) \nonumber  \\
    & - 8\pi\alpha\; W^2 S^{\rm TF}_{ij}\; ,\label{eq:bssn:conformal_traceless_extrinsic_curv} \\[2.5pt]
    \frac{d}{dt} \tilde{\Gamma}^i &= \tilde{\gamma}^{jk}\partial_j\partial_k\beta^i + \frac{1}{3}\tilde{\gamma}^{ij}\partial_k\partial_k\beta^k - 2\tilde{A}^{ij}\partial_j\alpha \nonumber \\
    & + 2\alpha\left(\tensor{\tilde{\Gamma}}{^i_{jk}}\tilde{A}^{jk} + 6\tilde{A}^{ij}\partial_j( \ln W^{-\frac{1}{2}}) -\frac{2}{3}\tilde{\gamma}^{ij}\partial_jK \right)\nonumber \\ 
    & - 16\pi\alpha W^{-2}j^i\; ,\label{eq:bssn:conformal_connection_function}
\end{align}
where $d/dt := \partial_t - \mathcal{L}_{\beta}$, expression $[\cdot]^{\rm TF}$ denotes the trace-free part of tensor inside the parenthesis, and $\tensor{\tilde{\Gamma}}{^i_{jk}}$ refer the Christoffel symbols with respect to $\tilde{\gamma}_{ij}$.
Additionally, the normal-normal and normal-parallel projections of the Einstein equations onto $\Sigma_t$ give us the Hamiltonian and momentum constraints, that written in the BSSN dynamical variables, are, respectively,
\begin{align}
    \label{eq:constr:hamiltonian}
    0 = \; H \; \equiv &\; R - \tilde{A}_{ij}\tilde{A}^{ij} + \frac{2}{3}K^2 - 16 \pi \rho\; ,\\
    \label{eq:constr:momentum}
    0 = M^i \equiv &\; \partial_j \tilde{A}^{ij} + \tilde{\Gamma}^i_{jk} \tilde{A}^{jk} + 6 \tilde{A}^{ij} \partial_j( \ln W^{-\frac{1}{2}}) - \frac{2}{3}\tilde{\gamma}^{ij}\partial_j K  \nonumber \\
    & - 8\pi W^{-2}j^i.
\end{align}

Concerning the evolution of the dynamical degrees of freedom of the Proca fields, it is known from the Hamiltonian formalism of electromagnetism that the physical dynamical variables of the system are $\{(a_m)_i,(\mathcal{E}_m)^i\}$, fulfilling evolution equations derived from the Hamilton equations. Moreover, the presence of $\Phi_m$ on the right-hand side of the evolution equation for $(a_m)_i$ (see Eq.~\eqref{eq:evol:vector_potential} below) compel us to also consider its  evolution equation derived from the Lorenz condition~\eqref{eq:Lorenz}. Thus, the evolution of the Proca fields will be set by the evolution of the system $\{\Phi_m,(a_m)_i,(\mathcal{E}_m)^i\}.$\footnote{{Since our formalism is based in the Hamiltonian formulation of Proca electromagnetism}, in this work we do not consider the ``magnetic'' field as a dynamical variable. Instead, we compute it every time using the identity derived from~\eqref{3+1:def:electricomagnetic_fields}, with the curl of the 3-vector potential, $(\mathcal{B}_m)_i = E^{ijk}D_j(a_m)_k$.} However, as this system contains non-propagating constraint violating modes~\cite{Zilhão:2015}, and even form an ill-posed Cauchy problem~\cite{Alcubierre:2009ij}, we follow the constraint damping procedure applied to the Maxwell electromagnetism as presented in~\cite{hilditch2013introduction}, where an additional dynamical variable $Z_m$ is introduced in the system through an evolution equation that sets its time derivative to be proportional to the Gauss constraint, and together with the remaining equations renders an extended system that is symmetric hyperbolic. Then, a term as $-\kappa Z_m$, with $\kappa>0$, is added to that evolution equation to damp this new field and hence the coupled constraint, recovering the original system for damped values of $Z_m$.\footnote{Analytically, the $Z_m$ should be zero to recover the original Proca system, but that is impossible due to the inherent discretization error of numerical methods.} 

In our context, the constraint of the system is given by the Gauss equation with a massive term, $0=D_i(\mathcal{E}_m)^i + m^2_V \Phi_m$, derived from the normal projection of~\eqref{eq:field:Proca}. Despite the fact that symmetric hyperbolicity was shown in~\cite{hilditch2013introduction} only for massless electromagnetism in curved spacetimes in the Lorenz gauge, the well-posedness can be safely extrapolated to the Proca system. This is because both systems share essentially the same characteristic structure since the additional massive terms in the evolution equations are of lower order and do not affect the hyperbolicity of the system. This formalism was successfully tested with numerical simulations of nearly-periodic real Proca fields around black holes in~\cite{Zilhão:2015}.
Therefore, the dynamics of our matter fields is determined by the evolution of $\{\Phi_m,(a_m)_i,(\mathcal{E}_m)^i,Z_m\}$. 
Their evolution equations read
\begin{align}
    \frac{d}{dt} \Phi_m \ &= \alpha\ [\ K\Phi_m  - Z_m  + \frac{1}{2}\Tilde{\gamma}^{ij}\partial_j\chi\ (a_m)_i \nonumber \\
    & - \chi \Tilde{\gamma}^{ij}\tilde{D}_j (a_m)_i \;] - \chi\Tilde{\gamma}^{ij}(a_m)_i \partial_j \alpha\; , \label{eq:evol:scalar_potential}\\[2.5pt]  
    \frac{d}{dt} (a_m)_i &= -\alpha\ [\ \chi^{-1}\Tilde{\gamma}_{ij}E^j + \partial_i \Phi_m \ ] - \Phi_m \partial_i \alpha \ , \label{eq:evol:vector_potential}\\[2.5pt]
    \frac{d}{dt} (\mathcal{E}_m)^i &= \alpha\ [ K(\mathcal{E}_m)^i + \chi \Tilde{\gamma}^{ij}\partial_j Z_m + m_V^2 \chi \Tilde{\gamma}^{ij}(a_m)_j   \nonumber  \\
    & + \frac{1}{2}\chi\Tilde{\gamma}^{ij}\Tilde{\gamma}^{pq}( \partial_j(a_m)_q \partial_p\chi - \partial_p(a_m)_j\partial_q\chi )\; ] \nonumber \\ 
    & + 2 \chi^2 \Tilde{\gamma}^{ij} \Tilde{\gamma}^{pq}( \alpha \Tilde{D}_p \partial_{[j}(a_m)_{q]} + \partial_q \alpha\:\partial_{[j}(a_m)_{p]})\:, \label{eq:evol:electric_field}\\
    \frac{d}{dt} Z_m \ \ &= + \alpha [\  \partial_i (\mathcal{E}_m)^i - \frac{3}{2}(\mathcal{E}_m)^i \partial_i(\ln{\chi)} + m_V^2 \Phi_m ]\   \nonumber \\ 
    &  - \alpha \kappa Z_m \; , \label{eq:evol:error_field}
\end{align}
where we use the conformal factor $\chi$ instead of $\phi$ as suggested by Campanelli \textit{et al.}~\cite{Campanelli:2005dd}, namely $\chi := e^{-4\phi}=W^2$. In the previous equation $\tilde{D}_i$ is the covariant derivative associated with the conformal metric. Notice also the presence of the Gauss constraint inside parentheses on the right hand side of Eq.~\eqref{eq:evol:error_field}.


\section{Numerical Setup}
\label{sec:numerical_setup} 

\subsection{Initial data}
\label{subsec:initial_data}

In order to perform a stability analysis, the configuration under study must first be perturbed to make it depart from equilibrium. In this work, the perturbations are mainly introduced as a result of the truncation error of our numerical calculations. Thus, we use the radial profiles $\{F_\ell(r), a_\ell(r), b_\ell(r), \alpha(r), \psi(r)\}$ and the time frequency $\omega$ of the equilibrium configurations that we already computed in~\cite{Lazarte:2024a} to obtain constraint-satisfying initial data for our evolutions. As the radial dependence of the orthogonal functions that describe our stars is computed using spectral projections, in order to import them onto the 3-dimensional Cartesian grid used in our evolution code 
we first evaluate the spectral solutions in a 1-dimensional grid, and then through first-order interpolations in isotropic coordinates~\eqref{metric:spherically-symmetryc}, we build the 3-metric $\gamma_{ij}$ and $\{ \Phi_m,\vec{a}_m,\vec{\mathcal{E}}_m\}$ in spherical coordinates at each grid point of the evolution code. The latter values, for the case $\ell=2$, describe the $2\ell+1=5$ complex Proca fields, with $m=\{-2,-1,0,1,2\}$, and are computed using the following expressions, obtained from applying~\eqref{3+1:def:potentials}-\eqref{3+1:def:electricomagnetic_fields} in \eqref{ansatz:potential:1-form} at $t=0$:
\begin{align}
    \label{ansatz:scalar_potential}
    \Phi_m &= (F_2/\alpha) \; Y^{2 m} \; , \\[5pt] \label{ansatz:vector_potential}
    \vec{a}_m &= ia_2\;Y^{2m}\hat{r} + ib_2\; \partial_\theta Y^{2m}\hat{\theta}+ ib_2\;\partial_\varphi Y^{2 m}\hat{\varphi} \; ,  \\[5pt]
    \label{ansatz:electric_field}
    \vec{\mathcal{E}}_m &= e_2\;Y^{2 m}\hat{r} + d_2\; \partial_\theta Y^{2 m}\hat{\theta}+ d_2\;\partial_\varphi Y^{2m}\hat{\varphi} \; ,
\end{align}
where we already explicitly write $\ell=2$ in the radial profiles. In particular, those for the ``electric'' fields were obtained as $e_\ell = -\left( F_\ell' + \omega a_\ell \right)/\alpha$ and $d_\ell = -\left( F_\ell  + \omega b_\ell \right)/\alpha$, with $'$ the radial derivative. Finally, the vector and tensor quantities in spherical coordinates are transformed into Cartesian coordinates. 

Regarding the initial data for the BSSN variables, $W$, $\tilde{\gamma}$, and $\tilde{\Gamma}^i$ are computed from $\gamma_{ij}$, and the rest are set to zero since the extrinsic curvature is initially $K_{ij}=0$ for a static spacetime. For the Proca system, the initial data for the fields $Z_m$ are also set to zero. To reduce interpolation errors, we interpolated the 1D-grid data with a resolution of about one-tenth of the highest resolution of the 3D-grid evolution code. 

\setlength{\tabcolsep}{6.5pt}
\begin{table}[h]
\centering
\begin{tabular}{c c c c c c }
\\ \hline
Model & $\alpha_0$ & $\omega/m_V$  & $m_V M_{\rm Komar}$  & $m_VR_{99}$ & $C_{99}$   \\ \hline\hline
$A$ & 0.620 & 0.794 & 2.006 & 11.632 & 0.172 \\
$B_1$ & 0.750 & 0.857 & 2.106 & 17.024 & 0.124 \\
$B_2$ & 0.760 & 0.859 & 2.101 & 17.576 & 0.120 \\
$C$ & 0.800 & 0.880 & 2.056 & 20.161 & 0.102 \\ 
$D$ & 0.840 & 0.903 & 1.967 & 23.541 & 0.084 \\ 
$E$ & 0.880 & 0.926 & 1.816 & 28.302 & 0.064 \\ \hline
\end{tabular}
\caption{Selected 2-Proca star initial configurations. The columns report the name of the model, the central value of the lapse, the time frequency $\omega/m_V$, and the star properties (mass, radius, and compactness; $m_VM_{\rm Komar}$, $m_VR_{99}$ and $C_{99}$). Details about the computation of the latter can be found in~\cite{Lazarte:2024a}.}
\label{table:2-Proca_configurations_initial}
\end{table}

\begin{figure}[b!]
\centering
\includegraphics[width=0.48\textwidth]{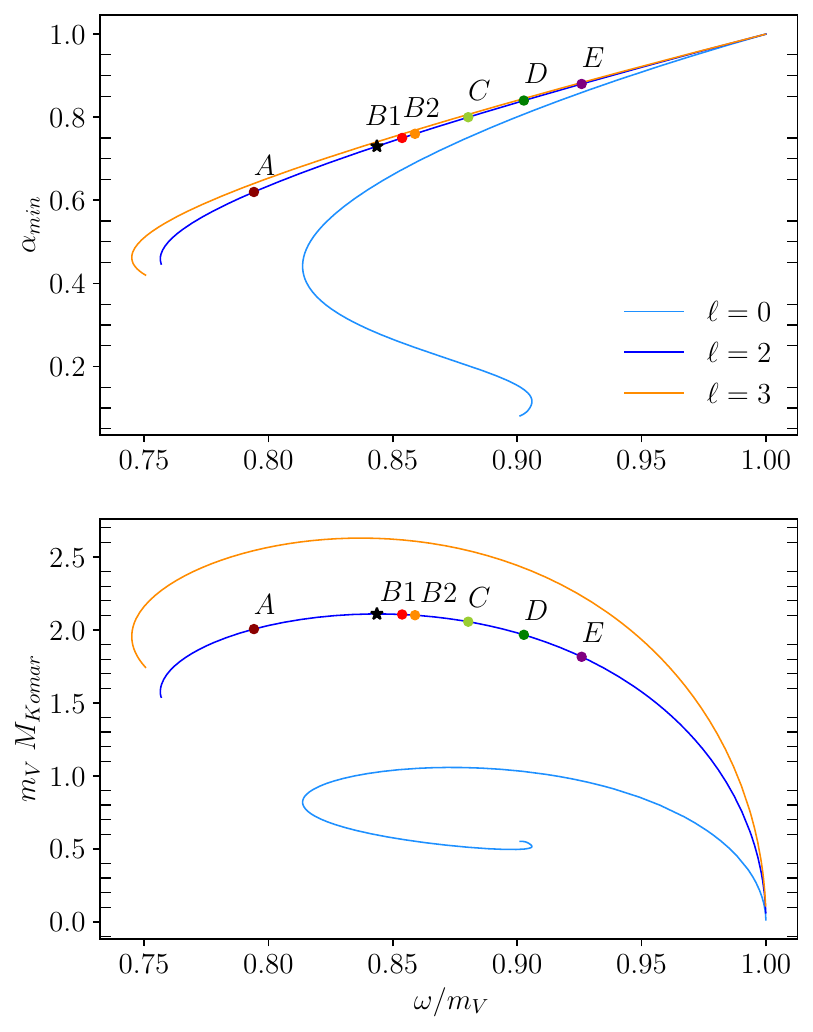}
\caption{Solution space of $\ell$-Proca stars for $\ell=0,2,3$. The specific configurations highlighted by dots correspond the ones specified in Table \ref{table:2-Proca_configurations_initial} in the domain of existence of the 2-Proca star. The black star corresponds the solution with maximum mass dividing this domain in two branches.}
\label{figure:solution_space}
\end{figure}

The domains of existence of the equilibrium solutions for $\ell$-Proca stars with $\ell=0,2,3$ 
are displayed in Fig.~\ref{figure:solution_space} through relations between the minimum lapse vs the frequency (top panel) and the Komar mass (see definition in Eq.~\eqref{eq:Komar_mass}) vs the frequency (bottom panel).  From among the family of solutions for the ($\ell=2$)-Proca star, hereafter referred simply as 2-Proca star, 
we select six illustrative solutions as initial configurations, which are represented as solid circles and whose properties are listed in Table~\ref{table:2-Proca_configurations_initial}. We focus our attention on studying the stability properties of configurations placed at the expected radially stable branch of solutions, which corresponds to the portion of the domain of existence curve located on the right of the maximum mass configuration, denoted by a star in Fig.~\ref{figure:solution_space}. For completeness, only one solution placed in the expected radially unstable branch was studied (model A).

\subsection{Time evolutions}\label{subsec:evolutions}


The 3D numerical-relativity simulations of our initial data were performed using the open-source Einstein Toolkit infrastructure~\cite{EinsteinToolkit:2024_11,loffler2012einstein} based on the Cactus framework~\cite{Goodale:2002a}. To evolve the spacetime variables, we employed the McLachlan thorn~\cite{Brown:2009dd}, using the BSSN system of equations, expressed as~\eqref{eq:bssn:conformal_factor_W}--\eqref{eq:bssn:conformal_connection_function}, with the ``1+log'' slicing condition for the lapse~\cite{Bona95} and the ``$\Gamma$-driver'' condition for the shift vector~\cite{alcubierre2003gauge}. To evolve 2-Proca star configurations, we extended the thorn used in~\cite{Sanchis_Gual_2019fae} to evolve two complex Proca fields to now evolve five complex Proca fields, integrating the system of equations given by~\eqref{eq:evol:scalar_potential}--\eqref{eq:evol:error_field}. This extended thorn was initially built from the Proca thorn, which, in turn, was employed to evolve one real Proca field in~\cite{Zilhão:2015} and is publicly available in the Canuda library~\cite{witek_2023_7791842}. 

We also used the fixed mesh refinement system of the Carpet driver~\cite{Schnetter:2003rb}, which allowed us to arrange the grids as nested cubes, where each cube is set to have the double resolution of the next cube that contains it. 
We used three nested cubes (or refinement levels) for the evolution of model $A$ and four for the remaining models. For the 3-refinement-level grid the nested cubes have semiedge values from the outermost to the innermost level of $m_V x_{\rm max}=m_V y_{\rm max}=m_V z_{\rm max}=\{98.4,\ 49.2,\ 24.6\}$. Pursuing  numerical convergence, we then changed to a 4-refinement level grid with spatial domains given by $m_V\{144,\ 72, \ 36,\ 18\}$ (see next subsection for the underlying reason for this change). With these domains, the spherical extension containing the 99$\%$ of the mass of most of our models, deduced from their $R_{99}$ values and reported in Table~\ref{table:2-Proca_configurations_initial}, is almost completely inside the innermost cube, with the exception of models $D$ and $E$. To reduce possible errors caused by the matter reflections with the innermost boundary for these two cases, we extended this cube, leading to a new overall spatial domain of $\{144,\ 72, \ 36,\ 25.2\}$. We also used this modified spatial domain in model $C$, finding negligible differences with its fiducial domain.

\begin{figure}[t]
\centering
\includegraphics[width=0.50
\textwidth]{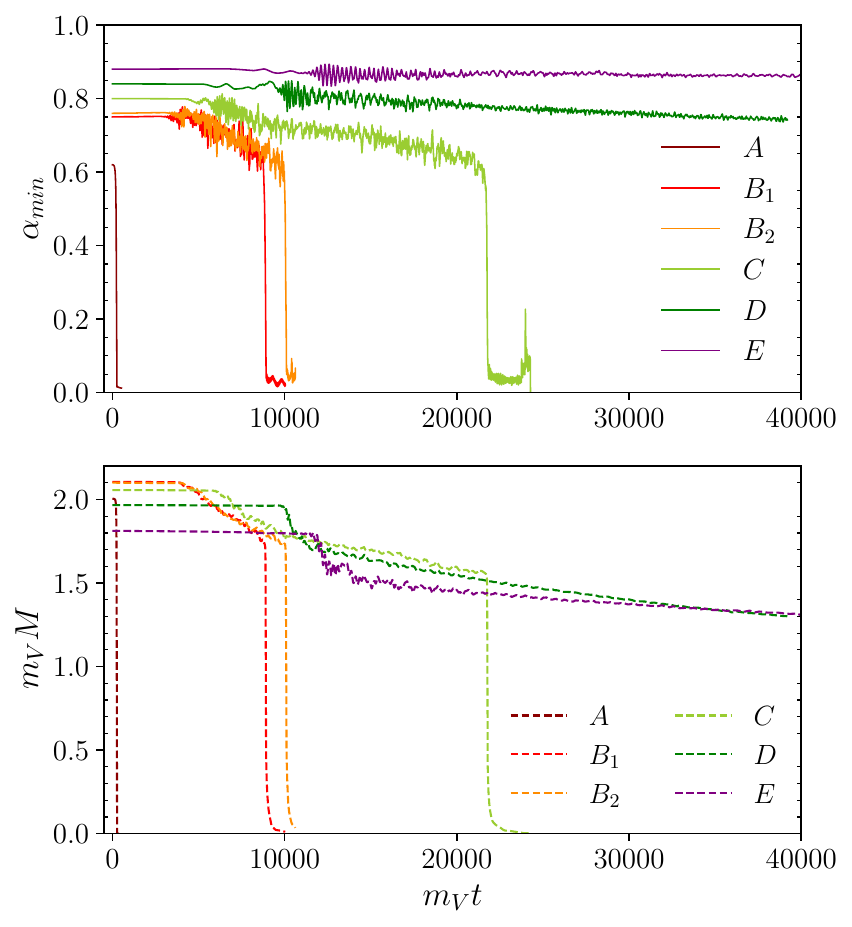}
\caption{Evolution of the minimum value of the lapse (top) and the 2-Proca star mass inside a sphere of radius $r_*=30$ (bottom) for the models $A$, $B_1$, $C$, and $D$ of Table~\ref{table:2-Proca_configurations_initial}.}
\label{fig:time-evolutions}
\end{figure}

The time integration was performed using the method of lines with a fourth-order accurate Runge-Kutta method, while the spatial derivatives were performed by centered fourth-order accurate first- and second-finite-differencing operators, except for the advection terms which are up-winded (also fourth order). Therefore, our evolution code is fully fourth-order accurate in time and space. All our simulations were performed using a Courant-Friedrichs-Lewy factor of $0.125$ and the fifth-order Kreiss-Oliger scheme with a dissipation parameter of $0.1$. We applied standard radiation (Sommerfeld) boundary conditions to all the evolved variables, including the falling power correction due to potential non-radiative degrees of freedom of our solutions~\cite{alcubierre2003gauge}. In particular, for the evolution of the Proca system variables we used a power of $n=3$ for scalar variables $\{ \Phi_m, Z_m\}$ and $n=2$ for the vector variables $\{ \vec{a}_m, \vec{\mathcal{E}}_m\}$.  

The total number of fields we have to evolve is fairly large. The matter fields comprise the real and imaginary components of each complex dynamical variable $\{ \Phi_m,\vec{a}_m,\vec{\mathcal{E}}_m, Z_m\}$, which implies 80 real variables describing the complete matter content of a 2-Proca star. Adding up the spacetime variables, we need to evolve more than 100 variables for each stellar model.

Figure~\ref{fig:time-evolutions} plots the evolution of the minimum value of the lapse function and the Komar mass of all $2$-Proca star models in our sample. As can be seen in Fig.~\ref{fig:time-evolutions}, in order to determine the end point of our simulations it was necessary to perform long-term evolutions. This makes our numerical simulations very computationally expensive, 
each taking several weeks of wall time to finish with resolutions within the convergence zone.


\subsection{Convergence tests}
\label{subsec:convergence}

\begin{figure}[h]
\centering
\includegraphics[width=0.49\textwidth]{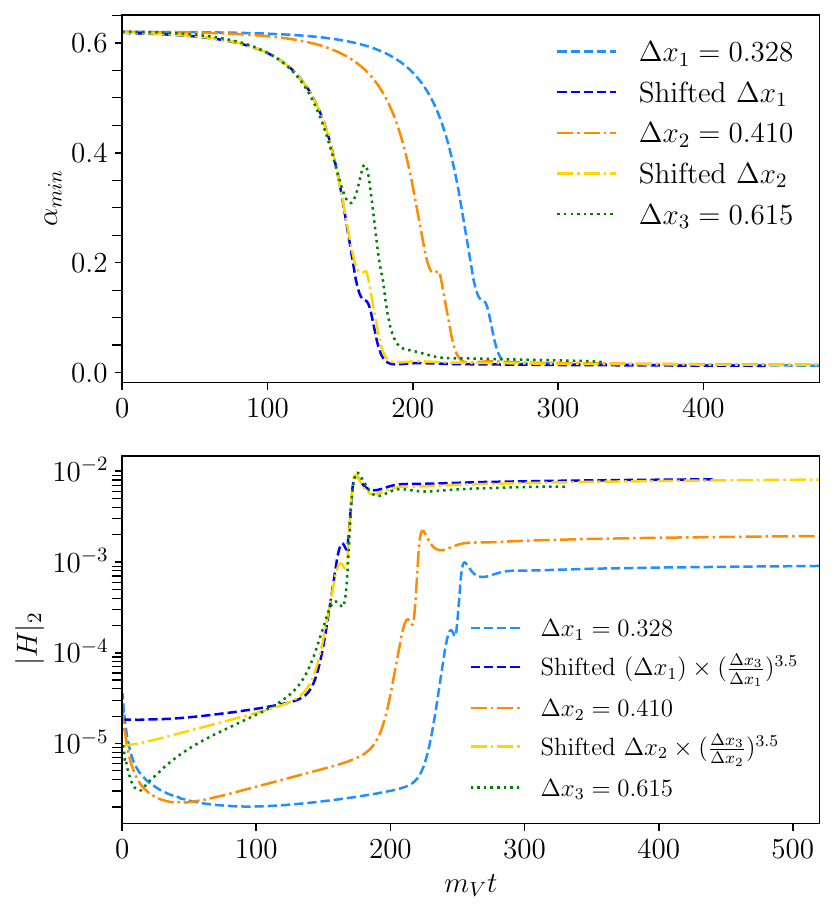}
\caption{Top: time evolution of the minimum value of the lapse for model $A$ and three different resolutions. Bottom: global convergence test for model $A$ using the $L_2$-norm of the Hamiltonian constraint. The curves are rescaled for an order of convergence between three and four.}
\label{fig:convergence_test_A}
\end{figure}

In order to determine whether our numerical results converge to physical solutions, we performed global and local convergence tests with three different resolutions, evaluating the evolution of the constraint violations of our numerical calculations. These tests also represent a validation of our extensions made to the thorn used to evolve the 2-Proca stars. 

{We restrict our analysis to perturbations arising from truncation errors for two main reasons: first, to avoid influencing the evolution of unstable modes when more than two are present in the configurations, as noted in~\cite{baiotti2007accurate}; second, for simplicity, although this implies that the perturbations may vary with resolution and could, in principle, lead to different late-time behaviors. Nonetheless, as we show in this section, our evolutions show the same physical behavior for different resolutions at late times and even convergence when appropriate time shifts are applied. These observations indicate that our perturbations at different resolutions differ mainly in amplitude while remaining qualitatively similar.}

Originally, we started with a global convergence test with the 3-refinement-level grid mentioned above, setting the low $m_V\{\Delta x, \Delta y, \Delta z \}=0.615$, medium $m_V\{\Delta x, \Delta y, \Delta z \}=0.410$, and high $m_V\{\Delta x, \Delta y, \Delta z \}=0.328$ resolutions of the finest (innermost) level. As a global test, we tracked the violations using the $L_2$-norm of the Hamiltonian constraint ($H$ defined in Eq.~\eqref{eq:constr:hamiltonian}), which is implemented in our code as $|H|_2:=(\;\sum^N_{i=0}H^2_i/N\;)^{1/2}$, with $N$ the total number of grid points. Figure~\ref{fig:convergence_test_A} shows the time evolutions of the minimum value of the lapse (top panel) and of $|H|_2$ (bottom panel) with these three resolutions for model $A$. The evolutions show that this configuration collapses and forms a BH for all three resolutions. The collapsing times are longer for finer resolutions. This is expected when truncation errors initially perturb the system and are the dominant errors throughout the simulation. Thus, at higher resolutions, the truncation error amplitudes are smaller and take more time to trigger gravitational collapse. Despite these time offsets, the minimum lapse evolution profiles are consistent with each other once a time shift is applied. On the other hand, the violation of the Hamiltonian constraint converges to zero by increasing the grid resolution with a convergence factor of about $3.5$, which is acceptable given the fourth-order accuracy of our evolution code. Notice that in order to see how the rescaled time evolution of $|H|_2$ overlay each other, we applied the same time shifts\footnote{{Our criterion for these time shifts was to remove the time offsets produced by the effect of different amplitude perturbations in the instability development, in such a way that the evolving quantities for different resolutions have the maximal overlap. This allows us to make comparisons between the evolution of the physical quantities and obtain the correct rescaling for the convergence test of evolutions dominated by truncation errors. The same criterion was applied, for instance, in~\cite{baiotti2007accurate} to compare the bar-mode instabilities seeded by truncation errors of rotating neutron stars in simulation with different resolutions.}} used for the minimum lapse evolutions. 
 
\begin{figure}[t]
\centering
\includegraphics[width=0.50\textwidth]{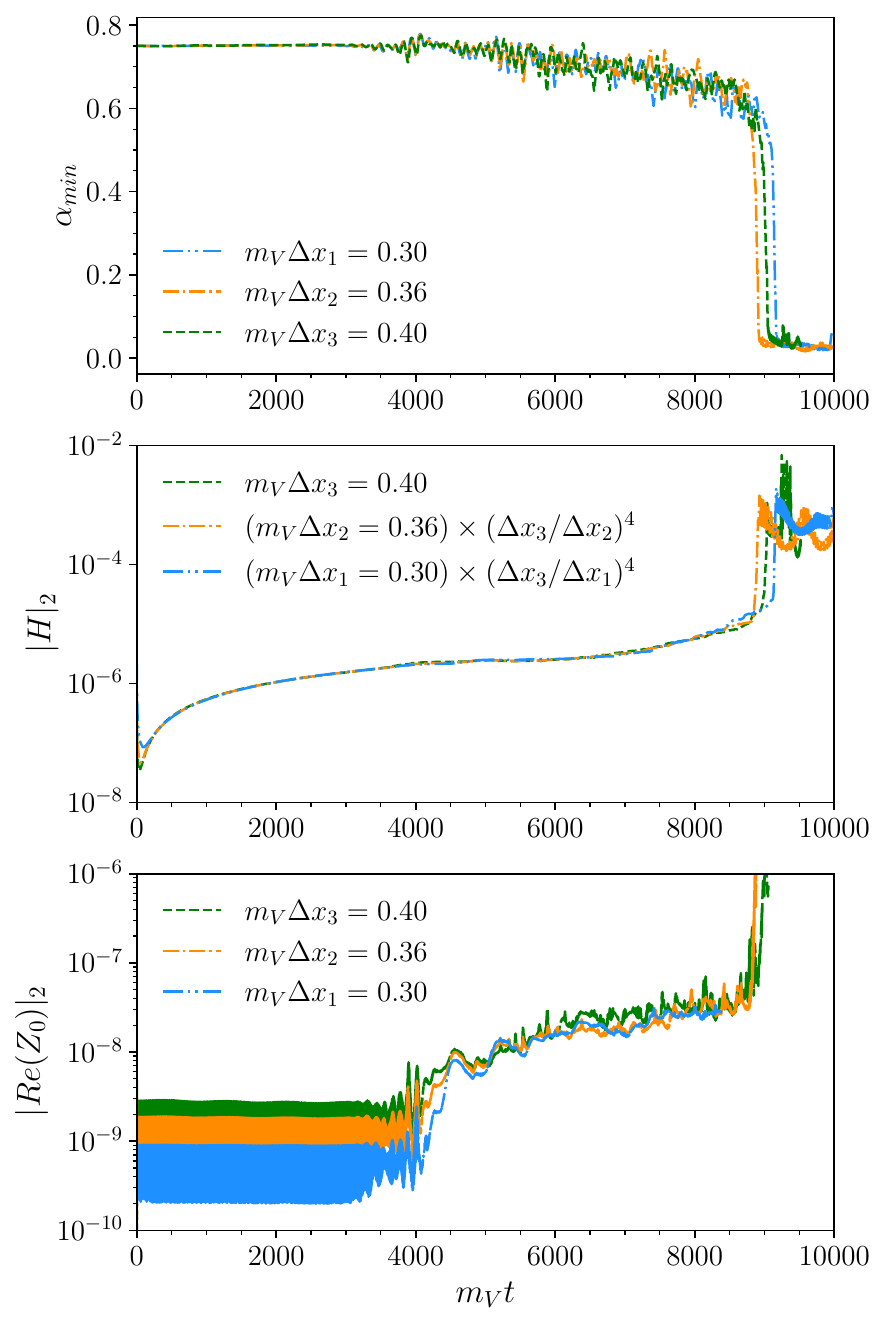}
\caption{Same as Fig.~\ref{fig:convergence_test_A} for model $B_1$ with a 4-refinement-level grid with the following resolutions for the innermost cube: low $m_V\{\Delta x, \Delta y, \Delta z\} =0.40 $, medium $m_V\{\Delta x, \Delta y, \Delta z\} =0.36$, and high $m_V\{\Delta x, \Delta y, \Delta z\} =0.30$, and including the $L_2$-norm of the Gauss constraint $Re(Z_0)$. {Unlike the model $A$, the timescale of the evolutions for this model is two orders of magnitude larger than the time offsets  caused by the different-resolution truncation errors, making them unnoticeable. In consequence, we do not apply the time shifts in these plots.}}
\label{figure:convergence_test_B1}
\end{figure}

We also performed a global convergence analysis of model $B_1$  using the previous 3-refinement-level grid but found inconsistent end-states across resolutions; namely, either the star collapsed to a BH at medium resolution or migrated to a seemingly stable configuration at low resolution. This lack of convergence indicated that our grid setup was insufficient to capture the correct dynamics. Furthermore, given that model $B_1$ lies within the expected radially stable branch, a collapse was unforeseen. 
After different tests, we observed that the loss of global fourth-order convergence coincided with the star's migration phase, during which significant matter ejection and gravitational potential deepening altered the star's behavior. These transitions, visible in the evolution of the Proca mass and the minimum lapse in Fig.~\ref{fig:time-evolutions}, led to more compact energy density profiles and irregularities in the Hamiltonian constraint evolution, indicating a need for higher resolution. 

\begin{figure*}[t!]
\centering
\includegraphics[width=1.0
\textwidth]{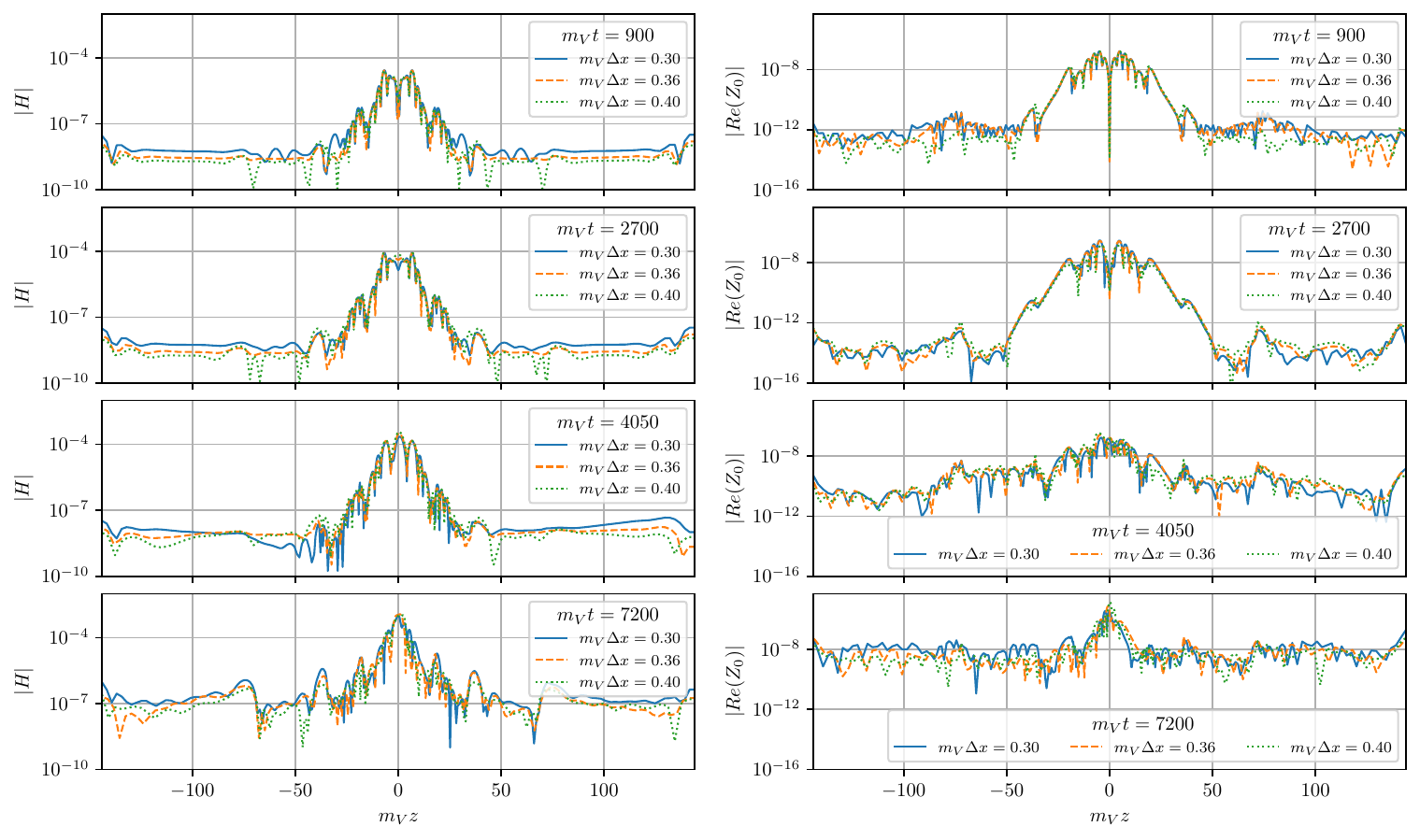}
\caption{Local convergence test for model $B_1$ using different snapshots of the evolution of $|H|$ (left) and $|Re(Z_0)|$ (right) and applying a fourth-order rate rescaling to the lower-resolution constraints.}
\label{fig:local_convergence}
\end{figure*}

To resolve this, we adopted a 4-refinement-level grid with an innermost domain smaller than our previous grid, allowing for accurate evolutions during the post-migration phase while limiting computational cost. Besides, an additional coarser outer level ensured sufficiently large radii for gravitational wave extraction. With this new setup, defined by spatial domains $ m_V x_{\max} = m_V y_{\max} = m_V z_{\max} = \{144,\ 72,\ 36,\ 18\} $, the convergence analysis showed collapse to a BH across all resolutions and a clear global fourth-order convergence throughout the evolution, as depicted respectively in the two first panels in Figure~\ref{figure:convergence_test_B1}. Since this model suffers from a migration of the Proca fields with a remnant vertical oscillation, we complemented our global convergence test by analyzing the $L_2$-norm of the Gauss constraint of the Proca field with $m=0$, which is precisely the field that when it migrates induces an oscillation of the whole system along the $z$ axis. Such a constraint is encoded by the variable $Re(Z_0)$, as we explained at the end of subsection~\ref{sec:3+1decomposition}. In the bottom panel of Figure~\ref{figure:convergence_test_B1}, we present these norms without rescaling for the three different resolutions, showing that the errors in this constraint were two to three orders of magnitude below those in $H$. Furthermore, we verified that they converge with a global fourth-order rate from the beginning until a time of $\sim 4500$, when they change their convergence rate to first order.   

To understand this decrease in the convergence ratio, we conducted a local convergence test of both previously analyzed constraints with the same resolutions used in the global test. Several snapshots of the evolution of the absolute value of $H$ and $Re(Z_0)$ are shown in Figure~\ref{fig:local_convergence}. In all shown times, the constraints for lower resolutions are rescaled using a fourth-order rate finding excellent overlaps between each other at the star's region and its surroundings, especially for the initial times. In particular, the local errors of $|Re(Z_0)|$ are very small and get even smaller in the outer levels thanks to the damping constraint technique used in our formalism; however, with the last two snapshots of this constraint we can verify that the floor error is increasing several orders of magnitude reaching the same orders around the star. This is consistent with a contamination of small local errors from matter reflections against the inner and outer boundaries propagating inward. Taking into account that we have a fixed grid, they may be further enhanced once the $\pi$-symmetry is broken due to the oscillation in the $z$ axis. These reflections do not follow physical equations, and as local effects, they accumulate for longer times contaminating the global quantities like the $L_2$-norms of the constraints. Thus, they first spoil the global convergence rate of small quantities as $|Re(Z_0)|_2$ and, then once they accumulate enough, could also affect the final stages of the $|H|_2$ evolution. In fact, this boundary reflection contamination plus the necessity of higher resolutions in the stage where the star collapses to BH can explain why the low-resolution simulation ($m_V\Delta x = 0.40$) leads to BH formation later than in the medium-resolution simulation. Therefore, our constraint violations converge at the expected rate globally at early times and locally throughout the evolution, especially in times when the star is facing their instabilities. We conclude that our setup achieves the convergence regime at higher resolutions, and we adopt the medium resolution ($m_V\Delta x = 0.36$) as the standard for the remaining models listed in Table~\ref{table:2-Proca_configurations_initial}. Regarding these cases and their longer timescales, we expect local errors to remain relatively small, since the models with end states different from $B_1$ ceased their vertical oscillations, and the two most dilute cases are well resolved as they do not collapse to a BH.

\section{Stability analysis}
\label{sec:stability_analysis}

\subsection{Methodology of the analysis}

Our stability analysis is performed by monitoring the dynamics of the Proca star throughout its evolution. The global dynamics is characterized with the minimum value of the lapse 
and with the Komar mass enclosed inside a sphere of radius $r_*$. The latter is computed from the following 3-volume integral~\cite{Gourgoulhon2012},
\begin{equation}\label{eq:Komar_mass}
    M_{r_*} = - \int^{r_*}_0 dr \int^{\pi}_0 d\theta \int^{2\pi}_0 d\varphi\; \alpha\sqrt{\gamma}\; (2\tensor{T}{^t_t}-T)\; ,
\end{equation}
where $T$ is the trace of the stress-energy tensor. Fig.~\ref{fig:time-evolutions} displays these two quantities for all our models, using an integration radius $r_*=30$. Notice that all stars have an initial effective radius $m_V R_{99}$ smaller than $r_*=30$ (see Table~\ref{table:2-Proca_configurations_initial}) and, when evolved, they migrate to more compact configurations with even smaller effective radii, ensuring they remain within the integration radius. Moreover, this choice ensures that any Proca field quantities radiated during the evolution are excluded from the integration. {We chose the Komar mass because our interest is in quasi-stationary final configurations, for which the Komar and Arnowitt-Deser-Misner masses coincide. In practice, after monitoring the minimum of the lapse function and discarding cases where black holes have linear momentum or form binaries, we found that the end states of our evolutions are consistent with quasi-stationary Proca solutions. In this regime, the Komar mass provides a reliable and straightforward measure of the system’s mass.} Furthermore, despite the fact that our system is initially static due to the mutual cancellation of the angular momentum between the field pairs $m=1$ and $m=-1$, and $m=2$ and $m=-2$, we also monitor the Komar angular momentum inside the same sphere of radius $r_*=30$, computed by~\cite{Gourgoulhon2012}
\begin{equation}\label{eq:Komar_angular_momentum}
    J_{r^*} = \int^{r_*}_0dr\int^{\pi}_0 d\theta \int^{\varphi}_0 d\varphi \; \alpha\sqrt{\gamma}\;\tensor{T}{^{\varphi}_t}\; .
\end{equation}

In~\cite{DiGiovanni_2020} it was shown that Proca stars with $m=|2|$ can develop bar-mode non-axisymmetric instabilities. To assess the possible onset of these instabilities,  we compute, as in~\cite{baiotti2007accurate,DiGiovanni_2020}, the azimuthal Fourier decomposition of the total energy density $\rho:=n^\mu n^\nu T_{\mu\nu}$ given by the following integration over the whole computational domain:
\begin{equation}\label{expr:azimuthal_modes}
   C_{\tilde{m}} :=  \int d \vec{x}^3 \sqrt{\gamma}\; \rho(\vec{x})\; e^{i\tilde{m}\tilde{\varphi}}\:.
\end{equation}
We stress that the azimuthal angle used in this expression is measured with respect to a system of cylindrical coordinates $(\tilde{r},\tilde{\varphi},\tilde{z})$ whose origin is located at the $Newtonian$ center of mass of the system measured with respect to the Cartesian grid coordinates\footnote{Although this is not a strictly Newtonian definition since the deformation of the spacetime is still present through a non-flat 3-metric and a non-trivial lapse, we refer to it as such because it is computed as the \textit{Newtonian} definition, namely, as the first momentum of the energy density divided by the total energy of the matter content.}, having its $\tilde{z}$ axis parallel to the $z$ axis of our Cartesian grid. Thus, $C_{\tilde{m}}$ quantifies {the distribution of total matter that presents a mode $\tilde{m}$} {around the $\bar{z}$ axis}. With these considerations, following~\cite{DiGiovanni_2020}, we take into account a possible emergence of a $\tilde{m}=1$ mode along the equatorial plane that could produce a non-zero linear momentum and subsequent kick displacing the center of mass.
In other words, the actual azimuthal angle used in the integration is $\tilde{\varphi}:=\arctan((y-y_{\rm CM})/(x-x_{\rm CM}))$. Therefore, we can extract from the $\tilde{m}$-mode $C_{\tilde{m}}$, its {mode} amplitude by evaluating the modulus $|C_{\tilde{m}}|$, and its {instantaneous} orientation by the phase $\tilde{\varphi}_{\tilde{m}}$ evaluated as
\begin{equation}
    \tilde{\varphi}_{\tilde{m}} = \arg(C_{\tilde{m}})\; . 
\end{equation}

\begin{figure}[t!]
\centering
\includegraphics[width=0.48
\textwidth]{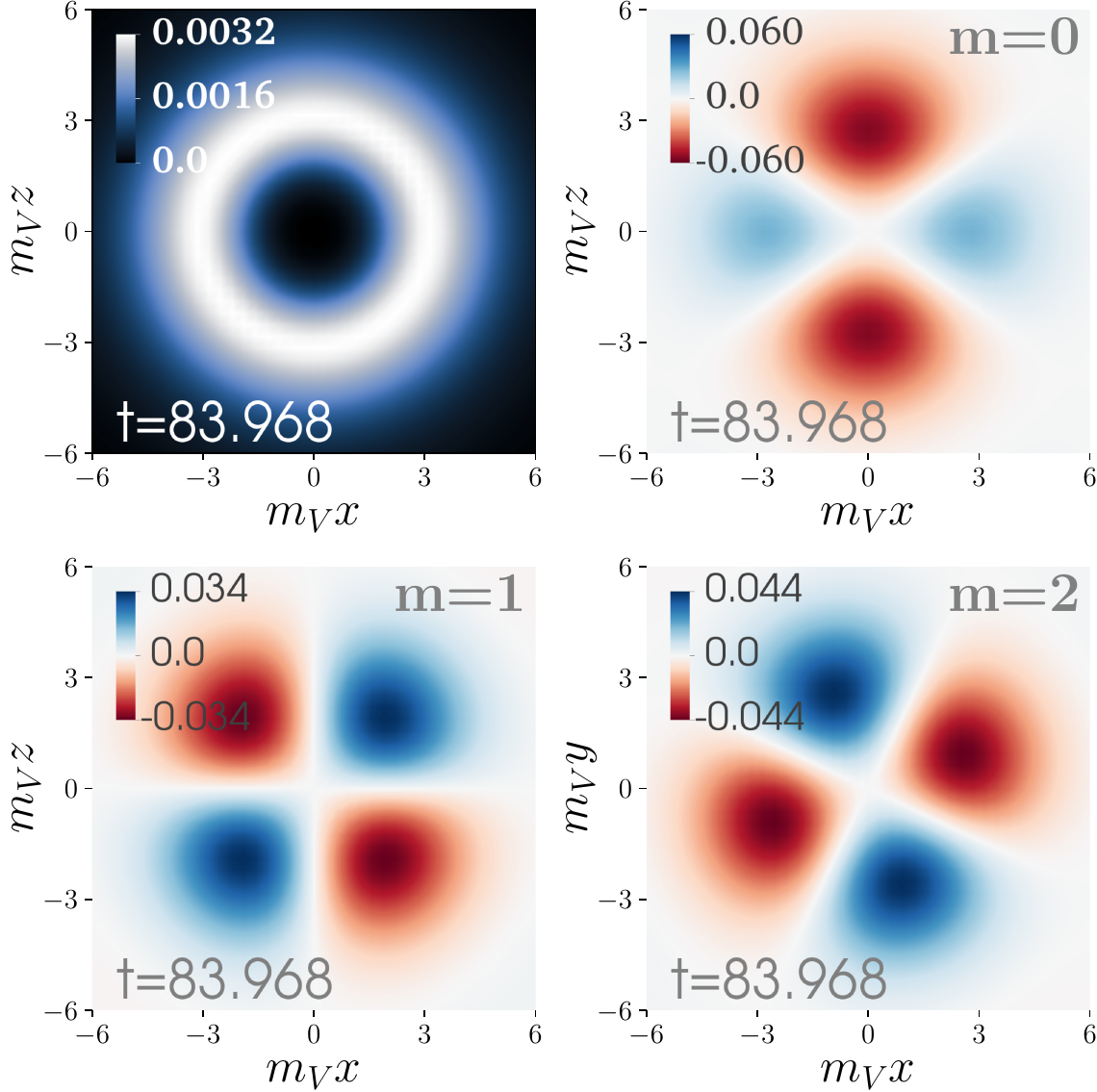}
\caption{2D snapshot (at $t=83.968$) of the total energy density in the $xz$ plane (top left) and ${\rm Re}(\Phi_m)$ of the constitutive fields of model $A$. 
The $m=0$ (top right) and $m=1$ (bottom left) fields are shown in the $xz$ plane, while the $m=2$ (bottom right) field is plotted in the $xy$ (equatorial) plane. The $m=-1$ and $m=-2$ fields, not shown here, have the same structure but rotated 45 degrees counterclockwise around the azimuthal axis (Cartesian grid $z$ axis). The total energy density displays the same spherical shell morphology in 3 plane views, $xy$, $yz$, and $xz$.}
\label{figure:A_2d}
\end{figure}

We also compute the gravitational wave emission through the evolution of our stars. This is done by considering the multipolar decomposition of outgoing gravitational waves computed through the Newman-Penrose (NP) scalar $\Psi_4$. This complex-valued scalar can be expressed asymptotically in terms of the gravitational wave amplitudes in the transverse-traceless gauge as 
\begin{equation}\label{psi_gw}
    \Psi_4 = -\ddot{h}_{+} + i \ddot{h}_{\times}\;,  
\end{equation}
where the double dot denotes the second time derivative and $h_{+}$ and $h_{\times}$ are the real-valued amplitudes of the plus and cross independent wave polarizations, respectively. The multipolar decomposition of the NP scalar is performed through projections onto the $-2$ spin-weighted spherical harmonic basis,
\begin{equation}\label{psi4_expansion}
    \Psi_4 = \sum_{\ell=0}^{\infty}\sum_{m=-\ell}^{\ell}\Psi^{\ell m}_4 (t,r) {_{-2}}Y^{\ell m}(\theta,\varphi)\; .
\end{equation}
The multipolar coefficients of $\Psi_4$ are useful to quantify the degree of symmetry of the matter distribution. According to the wave-generation formalism~\cite{Thorne80b}, these coefficients are connected to the mass $I^{\ell m}$ and current $S^{\ell m}$ multipoles of the gravitational radiation field as 
\begin{equation}
    \Psi^{\ell m}_4 (t,r) = \frac{1}{r\sqrt{2}}[\: {^{(\ell+2)}}I^{\ell m}(t-r) - i\: {^{(\ell+2)}}S^{\ell m}(t-r) ]\;,
\end{equation} 
where superscripts $(\ell+2)$ indicate the $(\ell+2)$th time derivatives and its dependence of the radial coordinate.  In this work we extract the multipolar coefficients at  different radii, namely $r = \{100,120,140\}$, verifying that these are within the local wave zone, and truncate the series in Eq.~\eqref{psi4_expansion} at term $\ell=4$. 

Finally, to provide a visual representation of the stellar dynamics, we have generated 2D and 3D snapshots of the energy density and the real part of the scalar potential of the constituent fields throughout the evolution.

\subsection{Loss of spherical symmetry and non-axisymmetric $\tilde{m}=4$ mode }

As explained in subsection~\ref{subsec:initial_data}, 2-Proca stars are composed of five complex independent Proca fields labeled with their spherical harmonic azimuthal number $m=-2,-1,0,1,2$, sharing the same harmonic time dependence (that is, the same $\omega/\mu_V$) and having a spatial dependence given by the ansatz~\eqref{ansatz:scalar_potential}--\eqref{ansatz:electric_field}. The result of their gravitational coupling is a static and spherically symmetric total stress-energy tensor. 

\begin{figure}[t!]
\centering
\includegraphics[width=0.5
\textwidth]{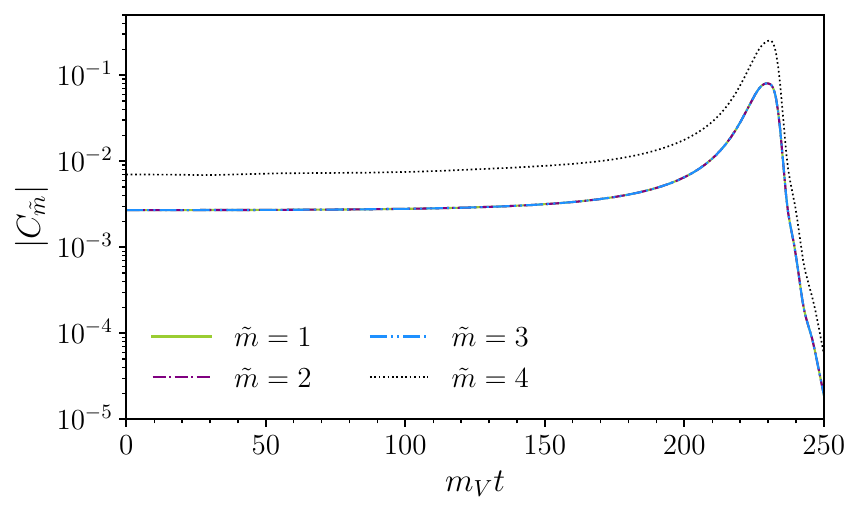}
\caption{Logarithmic-scale evolution of the azimuthal mode amplitudes for model $A$.}
\label{figure:A_aximuthal_modes}
\end{figure}

For model $A$, we verified the spatial spherical symmetry of the total energy density by examining its distribution on different planes, confirming that the spherical shell morphology is preserved throughout the entire evolution. Figure~\ref{figure:A_2d} shows 2D snapshots of the total energy density for this model (top left panel) and the real part of the scalar potentials of the constitutive fields with $m=0,1,$ and $2$. Even though the five fields evolve with different spatial dependence, the total energy density remains spherically symmetric. The evolution of this model indicates that the star collapses promptly into a BH as can be seen in the minimum value of the lapse and the mass of the star shown in Figure~\ref{fig:time-evolutions} (dark red curves). Additionally, Figure~\ref{figure:A_aximuthal_modes} reveals that the amplitudes of the $\tilde{m}$ modes have the same growth rates, indicating that the collapse is triggered by a radial instability.  We also observe that the $\tilde{m}=4$ mode has a slightly larger amplitude than the other azimuthal modes, an effect caused by the additional perturbation introduced by the Cartesian grid discretization. However, its growth rate follows the same trend as that of the other modes, which overlap one another throughout the evolution. Moreover, we do not find emission of gravitational radiation\footnote{In order to compute the multipoles of $\Psi_4$ for this model we performed an additional simulation with the 4-refinement-level grid with innermost level resolution $m_V dx= 0.36$ and with the domain extension described in  subsection~\ref{subsec:convergence}.}. The gravitational collapse observed for model $A$ is hence consistent with the expected outcome for a spherical 2-Proca star located in the unstable branch as a result of a radial instability. 


\begin{figure}[t!]
\centering
\includegraphics[width=0.5
\textwidth]{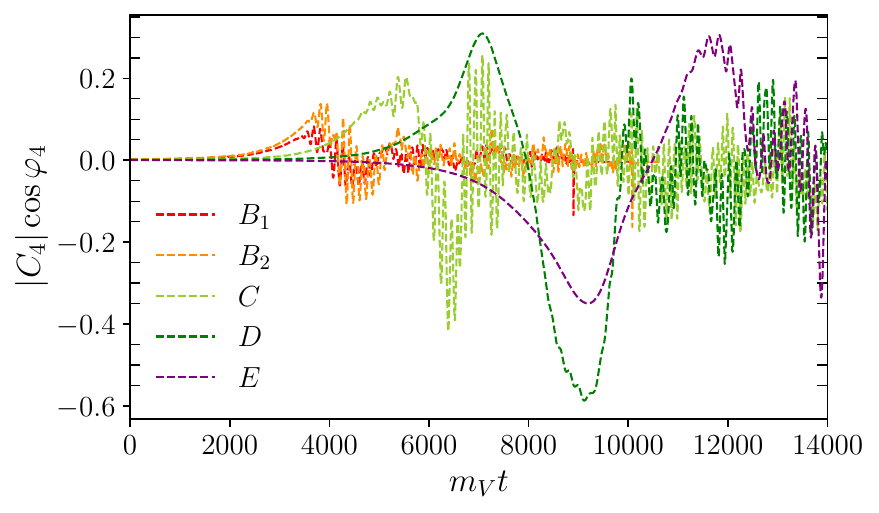}
\caption{Evolution of the real part of the azimuthal modes with $\tilde{m}=4$ for models $B_1$ to $E$ in linear scale, all with the same grid resolution.}
\label{figure:overal_4modes}
\end{figure}

Time evolutions of the remaining models are displayed in~Figure~\ref{fig:time-evolutions}. Initially these stars, located in the radially stable branch, conserve their spherical symmetry and are stable for significantly longer times than in the case of model $A$. The evolution of their $\tilde{m}=4$ azimuthal mode is plotted in Fig.~\ref{figure:overal_4modes}. This figure shows this mode grows exponentially for all models, revealing a distinct departure from spherical symmetry. It also shows that
the growth timescale of the $\tilde{m}=4$ mode depends on the compactness of the star.
We note that each mode is characterized by a complex number (real and imaginary part) that is set by its amplitude and its orientation around the $z$ axis, $C_{\tilde{m}}=(\;|C_{\tilde{m}}|\cos{{\varphi}_{\tilde{m}}}\;,|C_{\tilde{m}}|\sin{{\varphi}_{\tilde{m}}}\;)$. The $\tilde{m}=4$ mode is triggered by the perturbation introduced by the Cartesian grid discretization, growing in amplitude from the linear to the nonlinear regime, when it saturates and ceases to be a small perturbation to become part of the total distribution of matter. Although this mode is computed from the total energy density, its presence also affects the individual Proca fields, as we discuss below. Furthermore, both in the linear and nonlinear regimes this mode is nonrotating, unlike the cases analyzed in~\cite{DiGiovanni_2020,baiotti2007accurate}, since its imaginary part vanishes. Having a non-zero real part, the mode has a fixed orientation throughout the two regimes. The same non-spinning behavior is observed for the other $\tilde{m}$ modes. Therefore, although the initially spherical stars develop noticeable non-axisymmetric energy density distributions and oscillate radially once the modes saturate in the nonlinear regime, the constitutive fields still cancel the total angular momentum resulting in a non-rotating system. As a consistency check, we verified that the total angular momentum remains negligible during the evolution.

\begin{figure}[t!]
\centering
\includegraphics[width=0.5
\textwidth]{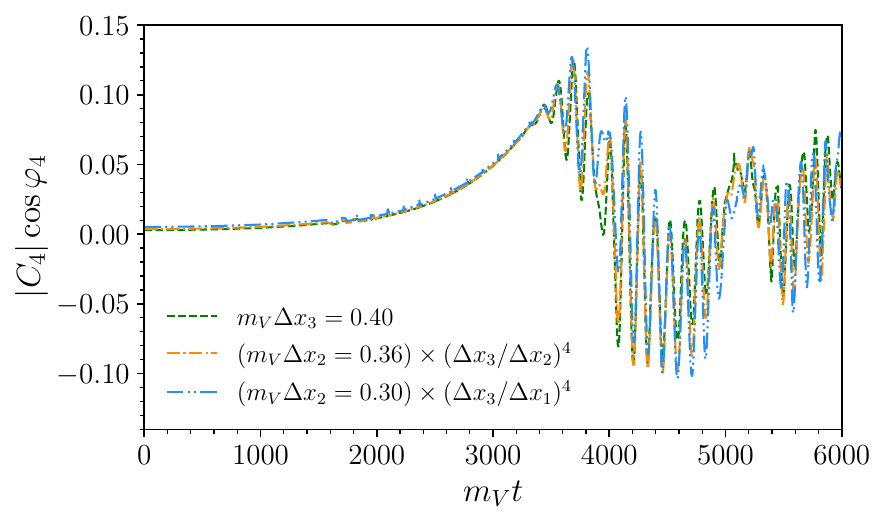}
\caption{Convergence test for the $\tilde{m}=4$ mode of model $B_1$ rescaled for fourth-order accuracy. The inset shows a magnification of the early evolution for different resolutions (indicated in the legend) with a second-order convergence at the initial time.}
\label{figure:B1_convergence_m4}
\end{figure}

Although one might be concerned by a possible numerical origin for the growth of the $\bar{m}=4$ mode, there exists previous evidence of its physical nature in studies of spinning scalar boson stars with nonlinear interactions~\cite{siemonsen2021stability} and static BH solutions~\cite{ridgway1995static}. In our case, we assess the physical nature of the $\tilde{m}=4$ mode by performing a convergence test on this quantity. This is shown in Figure~\ref{figure:B1_convergence_m4} for model $B_1$. We find that, at initial times, it converges to zero with second-order accuracy.{From the two initial left panels of Figure~\ref{fig:local_convergence}, we notice that $|H|$ converges well at fourth-order in the region of the star but not with that same order in the outer refinement levels. This local error is caused by the importation of the initial spectral data through interpolation. Hence, it is present in the integration over all the spatial space done to compute the azimuthal modes. However, numerical dissipation and the constraint damping of our numerical formalism ensure that the evolution quickly adjusts
itself to global fourth-order accuracy (the same error and self-adjustment was reported in~\cite{Pollney:2007ss})}. Therefore, the $\tilde{m}=4$ mode recovers the global fourth-order convergence in the linear regime of its exponential growth and even when it reaches the nonlinear phase, strongly indicating that this mode is physical. 

\subsection{Formation of multi-$\ell$ Proca stars}

\begin{figure}[t]
\centering
\includegraphics[width=0.48
\textwidth]{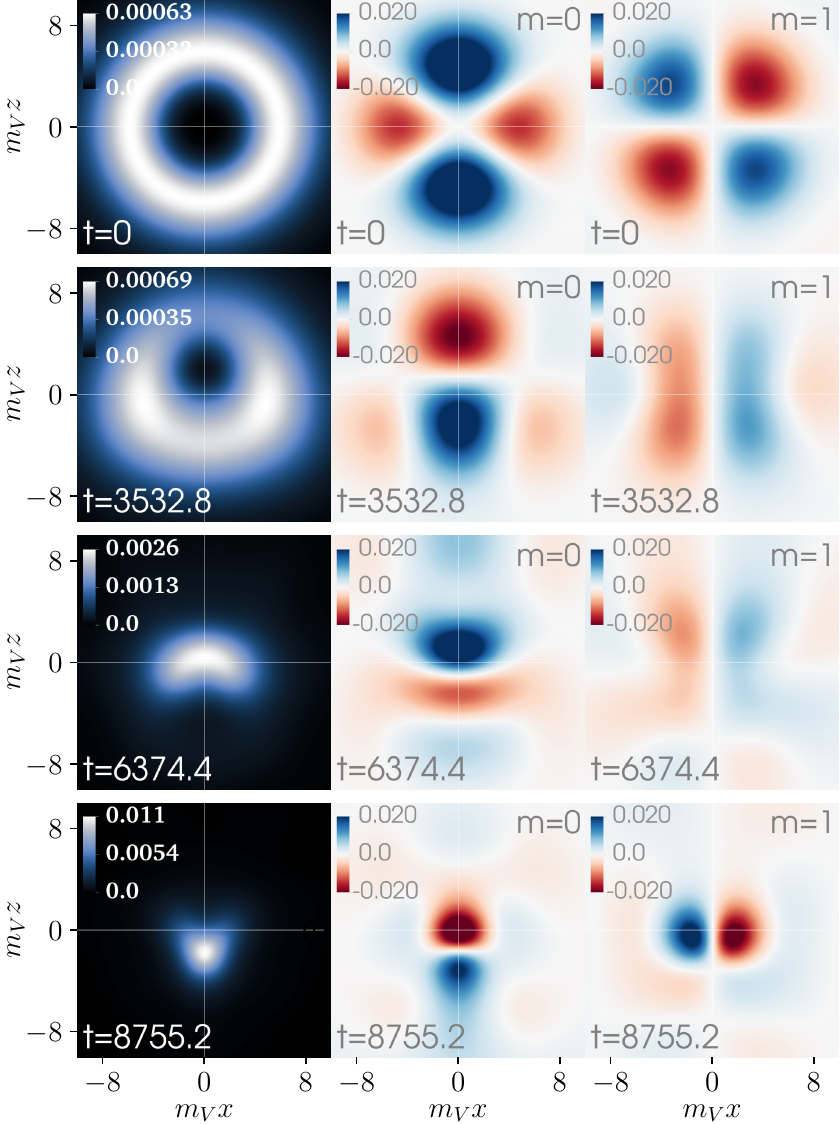}
\caption{Snapshots of the total energy density (left column) and ${\rm Re}(\Phi_m)$ of Proca fields with $m=0$ (central column) and $m=1$ (right column) in the $xz$ plane for model $B_1$. Model $B_2$ displays the same dynamical behavior.}
\label{figure:B1_2d_xz}
\end{figure}

\begin{figure}[t]
\centering
\includegraphics[width=0.48
\textwidth]{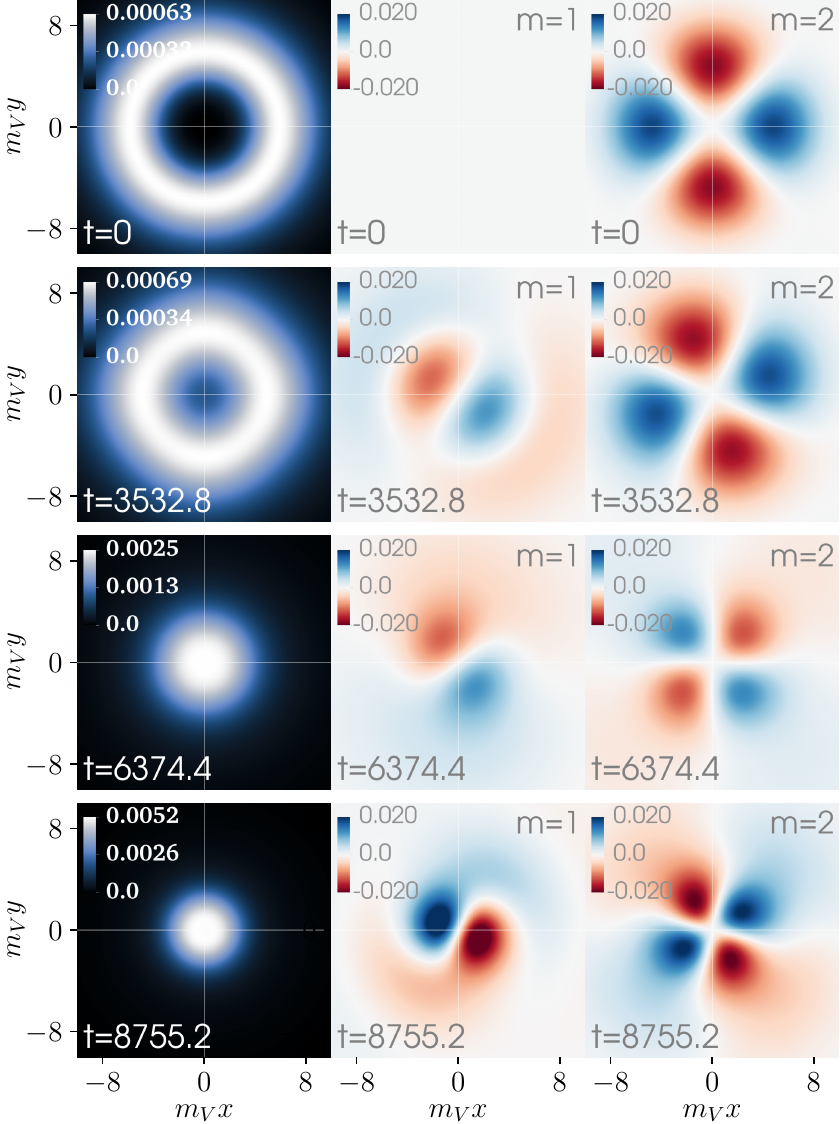}
\caption{Snapshots of the total energy density (left column) and ${\rm Re}(\Phi_m)$ of Proca field $m=1$ (central column) and $m=2$ (right column) in the $xy$ (equatorial) plane for model $B_1$. Model $B_2$ displays the same dynamical behavior.}
\label{figure:B1_2d_xy}
\end{figure}

When the non-axisymmetric $\tilde{m}=4$ instability reaches the nonlinear regime, it triggers the migration of the star to a new total energy density morphology, with a change in the angular momentum of some of the fields, and the emission of gravitational waves. The migration process of model $B_1$, which happens on a shorter timescale than for the rest of the models, is depicted in the first columns of Figs.~\ref{figure:B1_2d_xz} and~\ref{figure:B1_2d_xy}. Just as model $A$, the $B_1$ star is initially spherically symmetric. However, it does not collapse promptly to a BH. Instead, the evolution of this model is characterized by the appearance of an oscillatory deformation along the $z$-axis that breaks the equatorial symmetry. During this phase the system releases bosonic matter, as can be seen in the red curve in the bottom panel of Fig.~\ref{fig:time-evolutions} (reducing the mass of the star), and experiences a transformation into an object with spheroidal morphology. Throughout this process the energy density oscillates along the $z$-axis, a motion that persists until the end of the evolution (before the star collapses to a BH).
This process is depicted in the snapshots in the left column of Fig.~\ref{figure:B1_2d_xz}. We note that the maximum value of the energy density moves from the initial shell-like distribution to the final central location, as is also visible in the left column of Fig.~\ref{figure:B1_2d_xy}. 

The analysis of the evolution of the five Proca fields and of its real and imaginary components reveals that such a symmetry breaking is the result of the joint migration of the fields with $m=-1,0,1$ from the orbital momentum number $\ell=2$ to $\ell=1$, but still preserving their azimuthal number $m$. This 
occurs for all our models in the stable branch except for model $E$, which will be discussed in the next subsection. 
The migration of the fields can be seen in the snapshots depicted in the central column of Fig.~\ref{figure:B1_2d_xz},  showing the transformation of the non-rotating $(m=0)$-Proca field from a quadrupolar to a dipolar configuration (compare the top panel with the bottom panel). 
In addition, the right column of Fig.~\ref{figure:B1_2d_xz} and the central column of Fig.~\ref{figure:B1_2d_xy} display the rotating $(m=|1|)$-Proca fields at the $xz$ and $xy$ planes, respectively, showing a clear transition from quadrupolar to dipolar rotating configurations. 
The final dipoles are consistent with an $\ell=1$ angular momentum number. Such decay from $\ell=2\rightarrow\ell=1$ is supported by previous work. Indeed, the migration from quadrupolar ($\ell=2$, $m=0$) to dipolar ($\ell=1$, $m=0$) non-rotating Proca fields was presented in the evolution of multipolar axially symmetric Proca stars in~\cite{herdeiro2024non}. This reference showed that $(\ell=1$, $m=0)$-Proca stars are the ground state of the single-field non-rotating Proca star family, and thus there exist configurations that are  dynamically stable under generic non-linear perturbations. 
Our work presents for the first time the migration of the rotating $(\ell=2$, $m=|1|)$-Proca fields into $(\ell=1$, $m=|1|)$-Proca fields. This result is consistent with (and generalizes)  the findings reported in~\cite{Sanchis_Gual_2019fae,DiGiovanni_2020}, where it was shown that the ground state of the spinning single-field Proca star family is the configuration with $\ell=1$ and $m=|1|$, and with nodeless radial amplitudes in their scalar potential.
In~\cite{Sanchis_Gual_2019fae,DiGiovanni_2020} the migration of excited configurations with one node to the nodeless one was also reported. Therefore, our $(\ell=2$, $m=|1|)$-Proca fields begin their evolution in an excited state, having one radial node in its initial scalar potential~\cite{Lazarte:2024a}, and then migrate to the ground state configuration of their respective spinning family. (The difference with~\cite{Sanchis_Gual_2019fae, DiGiovanni_2020} is that our configuration has a higher angular momentum number.)  

Furthermore, the three fields with $\ell = 2,\ m = -1, 0, 1$ begin to decay simultaneously, strongly suggesting the formation of a new composite object emerging from part of the original system. Based on the energy distributions of the individual fields, the resulting configuration is expected to exhibit a spheroidal energy density morphology as an inherited feature from its constituent fields, which themselves display this morphology in their respective Proca star configurations, as reported in~\cite{Sanchis_Gual_2019fae,Sanchis-Gual:2021,herdeiro2024non}. 

It is worth stressing that if the $(m=|2|)$-fields were not present in the ($\ell=2$)-Proca star evolutions, the new composite star made of the remaining three fields could likely correspond to an axially symmetric ($\ell=1$)-Proca star configuration with high dynamical stability (a system we plan to study in the future). 
However, this is not our case, which renders useful taking into account the analysis of~\cite{DiGiovanni_2020} to understand the dynamics of Proca fields with $\ell=2$ and $m=|2|$ modes. In Ref.~\cite{DiGiovanni_2020} the fate of rotating $(m=|2|)$-Proca fields from constraint-satisfying initial data of dilute clouds was discussed, reporting the formation of a transient rotating $m=2$ Proca star with a toroidal energy density shape. After its formation, such configuration suffers a non-axisymmetric $\tilde{m}=2$ (bar-mode) instability that breaks its energy distribution into two pieces that subsequently merge into a $\ell=m=1$ rotating Proca star. 

Returning to our models, and taking the above considerations into account, we classify the stars into two categories based on their response against the development of the bar-mode instability. We will discuss the models that develop such an instability (models $C$, $D$, and $E$) in the next subsection and we focus here on those models ($B_1$ and $B_2$) that do not develop it. We shall emphasize that all five models share the evolution discussed above for model $B_1$ wherein the ($m=|2|$)-Proca fields keep the $\ell=2$, $m=|2|$ multipolar structure while the other fields migrate from an $\ell=2$ to an $\ell=1$ multipolar structure (cf.~right column of Fig.~\ref{figure:B1_2d_xy}).

\begin{figure}[t!]
\centering
\includegraphics[width=0.48
\textwidth]{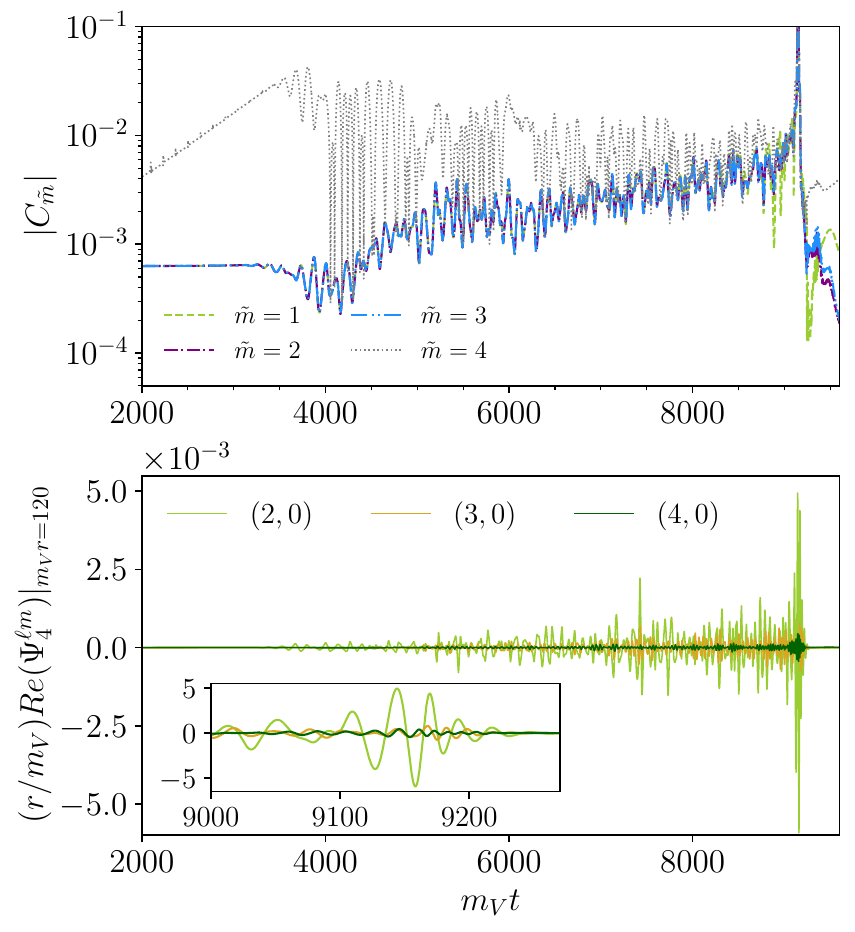}
\caption{Top: evolution of the azimuthal modes of model $B_1$. Bottom: dominant gravitational-wave modes for model $B_1$ extracted at radius $m_V r=120$. The maximum emission is produced at the moment of BH formation.}
\label{figure:B1_azimuthal_modes-gws}
\end{figure}

Our discussion so far has shown the formation of a new evolution stage where only some of the constitutive Proca fields of the 2-Proca stars migrate from a quadrupolar to a dipolar structure. In this new configuration the two different angular momentum numbers, $\ell=1$ and $\ell=2$, coexist. This configuration was already studied in the scalar sector of multi-field boson stars in~\cite{Alcubierre:2023}. Therefore, we recognize it as a multi-$\ell$ Proca star. The total energy density of the new configuration regime has a spheroidal shape with axial symmetry (cf.~Figs.~\ref{figure:B1_2d_xz} and~\ref{figure:B1_2d_xy}) in spite of the toroidal shape contribution in the energy density given by the non-migrating ($m=|2|$)-fields. At this stage, the $\tilde{m}=4$ perturbation still dominates over the other $\tilde{m}=1,2,3$ modes, as shown in the top panel of Fig.~\ref{figure:B1_azimuthal_modes-gws} for model $B_1$. However, these start to be excited due to the radial oscillations and have a fixed orientation which indicates that they are also non-rotating. These three azimuthal modes overlap with one another during the entire evolution, suggesting that the radial perturbations dominate the dynamics. Eventually they reach an amplitude comparable to that of the $\tilde{m}=4$ mode at which point the configuration collapses into a BH. Models $B_1$ and $B_2$ not only end up collapsing to form a BH but the latter suffers a kick due to the global dynamics of the system and the linear momentum acquired during the previous stages (see the red and orange curves in Fig.~\ref{fig:time-evolutions}).

When the models are in the multi-$\ell$ regime, the relevant multipolar modes of the gravitational-wave emission are  axisymmetric, namely $(\ell,m)=(2,0), (3,0), (4,0)$, dominated by the (2,0) quadrupolar mode with values about one order of magnitude larger than the others. This is shown for model $B_1$ in the bottom panel of Fig.~\ref{figure:B1_azimuthal_modes-gws}. 
All other available gravitational-wave modes are negligible since they acquire values several orders of magnitude smaller than the dominant (2,0) mode. Additionally, the axisymmetric modes $\Psi^{\ell,0}_4$ have a zero imaginary part. Thus, using equations~\eqref{psi_gw} and~\eqref{psi4_expansion}, we deduce that the gravitational waves emitted in this regime only have + polarization, proving the high degree of axial symmetry of the spacetime and subsequently of the matter source. This is a strong indication that our multi-$\ell$ Proca stars, despite having a hexadecapolar inner distribution of matter contaminating the system, have a total energy distribution which is approximately axially symmetric. 

The above analysis leads to the following interpretation of the evolution and end-point of models $B_1$ and $B_2$. First of all, $\ell$-Proca stars are multi-field extensions of spherical Proca stars. Those were initially thought to be stable under generic perturbations due to its dynamical stability under spherical perturbations in 1D-simulations. However, those are actually excited states, as it was proved in~\cite{herdeiro2024non} through the evolution of spherical Proca stars in 3D-simulations and without imposing any symmetry. Radially stable configurations migrate to ground state configurations characterized by a dipolar structure in their Proca potential (see evolutions of solutions 5 and 8 in Fig.~1 and Table 1 in~\cite{herdeiro2024non}), and they do not experience a radial collapse. This, in principle, seems in contradiction with our result for $B_1$ located in the expected stable branch against spherical perturbations. Nevertheless, we should not extrapolate directly the dynamics of ($\ell=0$)-Proca star to our ($\ell=2$)-Proca star. Instead, we must carefully examine the dynamical properties of their constitutive fields. In this sense, we notice that there are solutions of ($\ell=2$, $m=0$)-Proca stars placed in the radially stable branch (see solutions 11 and 12 in~\cite{herdeiro2024non}) that migrate into radially unstable $\ell=1$, $m=0$ configurations, subsequently collapsing into a BH~\cite{Herdeiro2023ProcaStarEvolution}. These solutions are initially placed close to the maximum mass solution as our model $B_1$. Therefore, the migration of our $B_1$ star to a radially unstable multi-$\ell$ Proca star is consistent with the dynamical instability of its $(m=0)$-Proca field. The robustness of this result was assessed with the evolution of model $B_2$, with parameters slightly different from those of model $B_1$ (cf.~Table~\ref{table:2-Proca_configurations_initial}). 
Model $B_2$ indeed faces exactly the same kind of evolution of its constitutive fields, forming a radially unstable multi-$\ell$ Proca star and collapsing to a BH within the dynamical time of our simulation (cf.~Fig.~\ref{fig:time-evolutions}).   

\subsection{Non-axisymmetric $\tilde{m}=2$  instability in multi-$\ell$ Proca stars} 

\begin{figure}[t!]
\centering
\includegraphics[width=0.48
\textwidth]{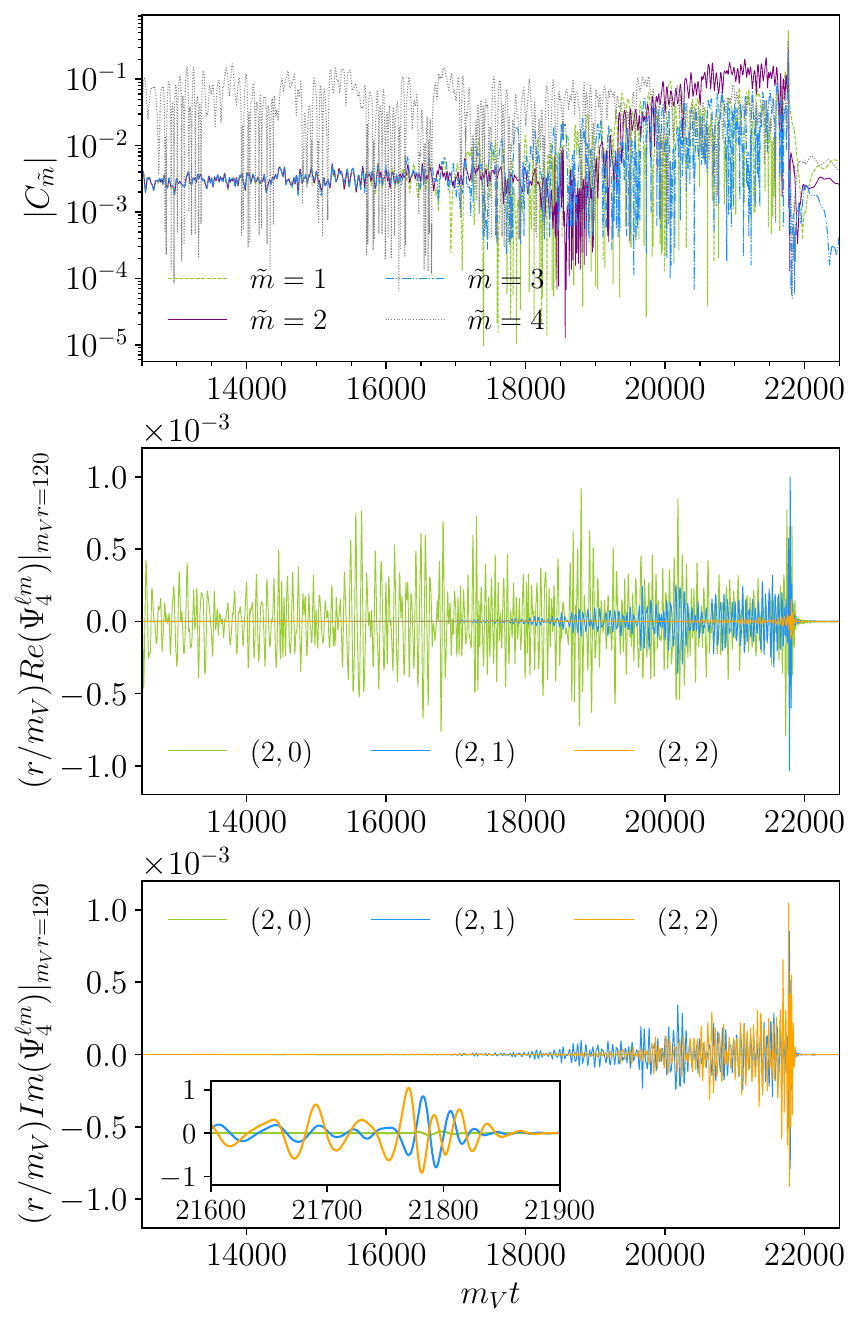}
\caption{Top: evolution of the azimuthal modes of model $C$. Middle and bottom: real and imaginary parts of the dominant GW modes for model $C$, respectively. All modes were extracted at radius $m_V r=120$. }
\label{figure:C_azimuthal_modes-gws}
\end{figure}

As discussed in the previous section, the evolutions of models $C$, $D$, and $E$ are characterized by the appearance of a bar-mode instability. 
Since these models do not radially collapse when they reach the multi-$\ell$ regime, we argue that they belong to the radially stable branch of multi-$\ell$ Proca stars. Although they are radially stable, we evolve them long enough to allow for the onset of a $\tilde{m}=2$ non-axisymmetric instability coming from the $m=|2|$-Proca fields. The outcome of this instability depends on the compactness of model. For the more compact model $C$, the end-product is a non-axisymmetric collapse to a BH as shown in the top panel of Fig.~\ref{figure:C_azimuthal_modes-gws}. This figure shows how the $\tilde{m}=1, 2, 3$ modes, which evolve identically up to $m_Vt\approx 17000$, begin to diverge and grow at different rates, reaching amplitudes comparable to that of the $\tilde{m}=4$ mode. Among them, the $\tilde{m}=2$ mode is the dominant one, with slightly higher values than the rest of the azimuthal modes. The emergence and dominance of this bar-mode instability, unlike for models $B_1$ and $B_2$, is responsible for the emission of non-axisymmetric gravitational-wave $m\neq0$ modes with non-zero real and imaginary parts. This is shown in the middle and bottom panel of Fig.~\ref{figure:C_azimuthal_modes-gws}. 

\begin{figure}[t!]
\centering
\includegraphics[width=0.48
\textwidth]{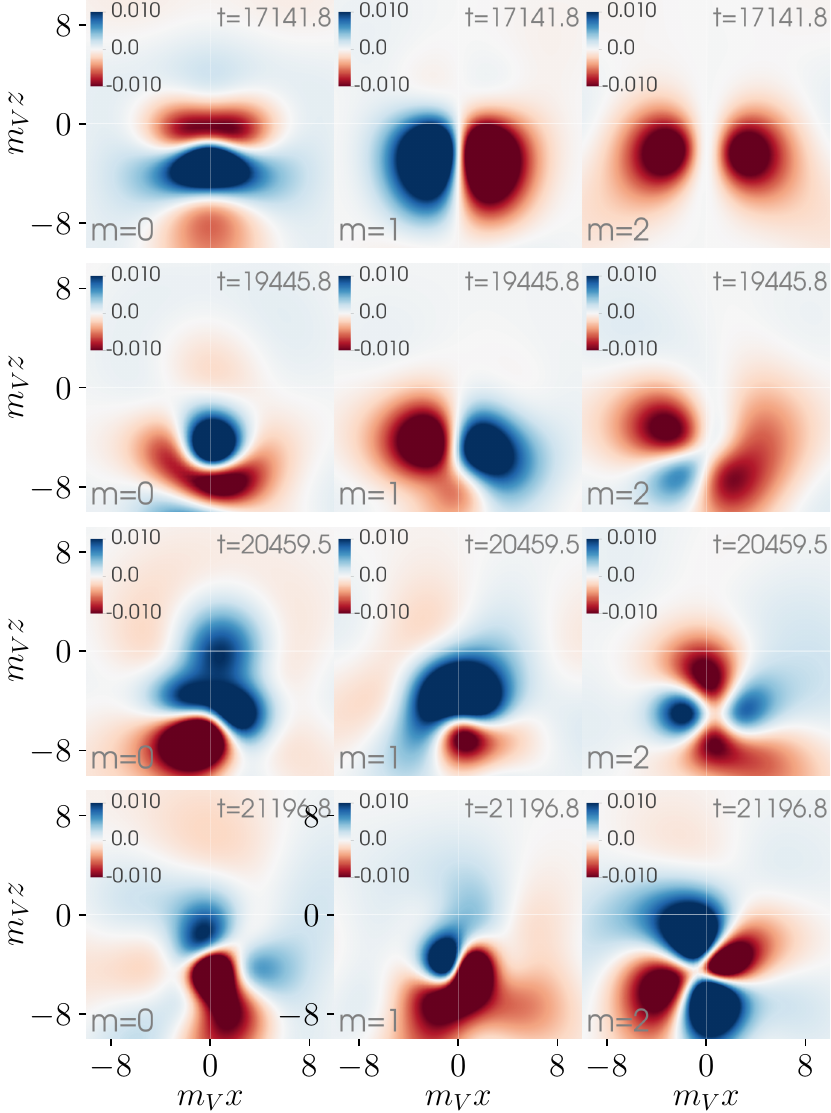}
\caption{Time evolution in the $xz$ plane cut of $Re(\Phi_m)$ of the constituent Proca fields with $m=0$ (left column), $m=1$ (central column) and $m=2$ (right column) for the model $C$.}
\label{figure:C_2d_xz}
\end{figure}

Figure~\ref{figure:C_2d_xz} shows 2D-snapshots in the $xz$ plane of the real part of the scalar potential for model $C$. The times selected correspond to the growth of the bar-mode instability. Once the $\tilde{m}=1,2,3$ modes begin to grow, there is a clear displacement of the internal center of all Proca fields. This is due to the joint effect of the excitation of the $\tilde{m}=1$ mode, introducing a linear momentum orthogonal to the $z$ axis, and the suppression  of the oscillatory movement of the star along that axis. The last two snapshots of Fig.~\ref{figure:C_2d_xz} show that the center of the Cartesian grid no longer coincides with the internal center of the Proca fields. This center is still the same for the five constitutive fields, showing that the star moves as a whole. Furthermore, the excitation of the odd $\tilde{m}$-modes likely creates asymmetries in the matter distribution that cause a somewhat erratic misalignment of the internal rotational axis of the spinning Proca fields with respect to the $z$-axis. This affects the rotation of the constitutive fields and can be especially observed as the change in the rotational plane of the ($m=|2|$) fields in the right column of Figure~\ref{figure:C_2d_xz}. 

\begin{figure}[t!]
\centering
\includegraphics[width=0.48
\textwidth]{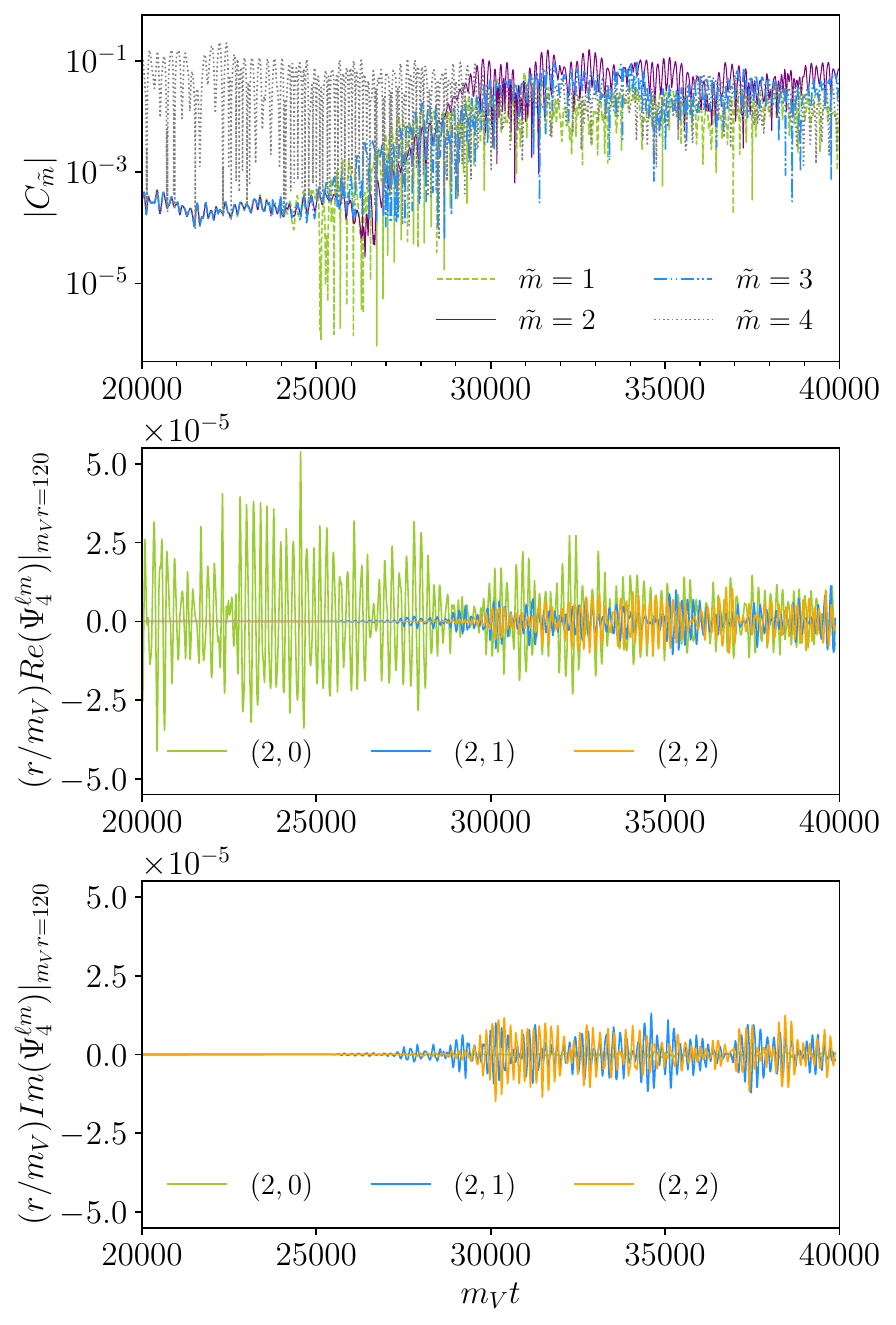}
\caption{As Fig.~\ref{figure:C_azimuthal_modes-gws} but for model $E$.}
\label{figure:E_azimuthal_modes-gws}
\end{figure}

\begin{figure*}[t!]
\centering
\includegraphics[width=0.95
\textwidth]{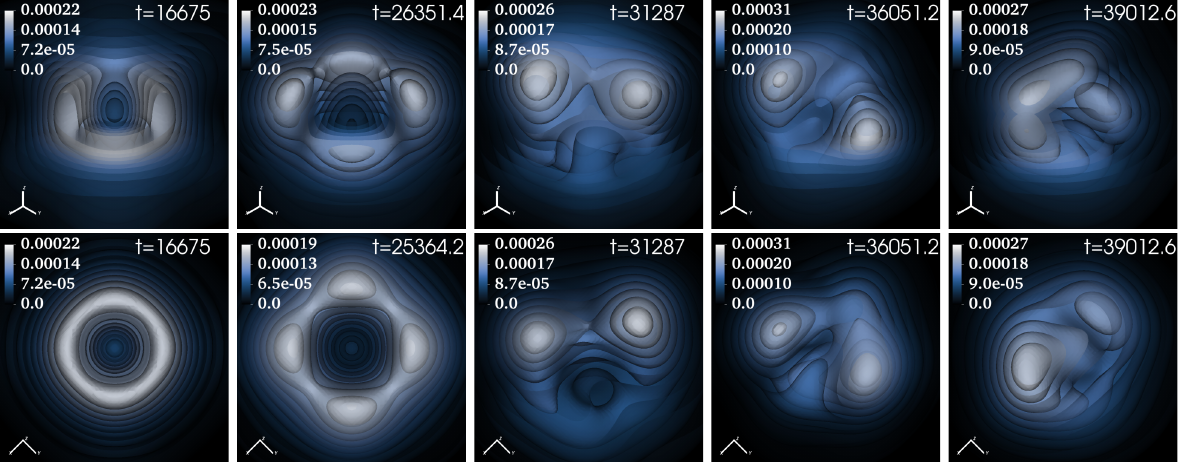}
\caption{Snapshots of the energy density of model $E$ created as a set of 3D iso-surfaces with a decreased opacity. A clipping operation was applied to remove appropriate sections, allowing for a clear visualization of the internal morphology. Top row: $(x,y,z)=(1,1,1)$ diagonal view without an octant of the total volume. Bottom row: equatorial plane view without the superior half of the volume.}
\label{fig:E_energy_density}
\end{figure*}

The collapse of model $C$ to a BH happens shortly after the $\tilde{m}=2$ mode saturates. This leaves little room to analyze the effects of the bar-mode instability in the energy density of this model\footnote{This was however possible in~\cite{DiGiovanni_2020} for single-field spinning Proca stars.}. However, this can be studied for the less compact model $E$ which does not collapse to a BH even for a significantly long evolution reaching up to $m_Vt=40000$ (see Fig.~\ref{fig:time-evolutions}). The evolution of its azimuthal modes and the real and imaginary parts of the dominant GW modes are displayed in Fig.~\ref{figure:E_azimuthal_modes-gws}. For model $E$, the bar-mode instability is strong enough to produce a fragmentation of the matter content of model $E$ in two individual clumps as shown in the 3D snapshots of Fig.~\ref{fig:E_energy_density}. The masses of the two clumps are not equal which we argue is due to the breaking of symmetries produced by the highly excited odd $\tilde{m}$-modes present during the last stages of the evolution, as we discussed for model $C$. The first row of Fig.~\ref{fig:E_energy_density} displays 3D snapshots that were rendered from a viewpoint aligned with the diagonal of the grid, along the $(x,y,z)=(1,1,1)$ direction. The second row corresponds to a viewpoint aligned with the positive $z$-axis, with the $xy$ plane facing the viewer. To facilitate the visualization of the internal features, a section corresponding to either one octant or half of the volume was removed in the first and second row, respectively. Notice that in the first snapshots, before the fragmentation process, the star does not have an actual spheroidal morphology, which would correspond to the multi-$\ell$ Proca star. Instead, the hexadecapolar structure developed in the first evolution phase is strongly present and critically shapes the energy density morphology. We attribute this to the fact that the migration of the $m=|1|$ fields is affected by a mode-mixing between its azimuthal dependency and the $\tilde{m}=4$ mode, since the amplitude of the latter is very large for the less compact configurations, as we observed in Fig.~\ref{figure:overal_4modes}. We checked that the $m=|1|$ Proca field indeed acquires a multipolar structure consistent with $m'=|3|$, coming from a mode-mixing of $m'=m\pm4$. In turn, this mixed multipolar structure forces the energy density to become toroidal with a square shape instead of fully spheroidal. 


\begin{figure*}[t]
\centering
\includegraphics[width=0.95
\textwidth]{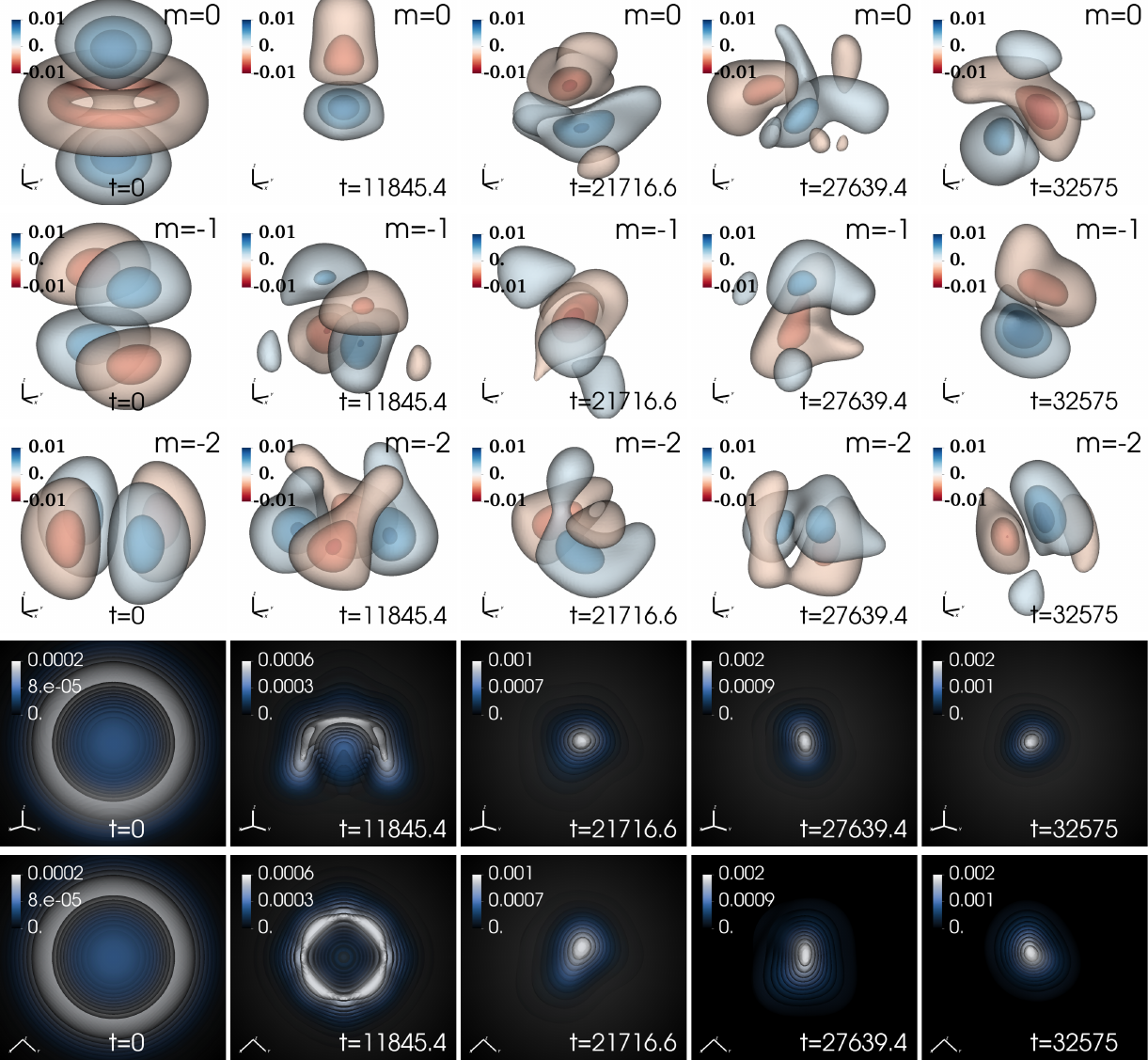}
\caption{Snapshots of the real part of ${\rm Re}(\Phi_m)$ for $m=0$ (first row), $m=-1$ (second row) and $m=-2$ (third row) and of the energy density (two last rows) of model $D$ created as a set of 3D iso-surfaces with a decreased opacity. The snapshots of the scalar potentials are created with the same fixed view. For the energy density we applied a clipping operation to remove half of the volume in front of the viewer. In particular, the fourth row uses a $(x,y,z)=(1,1,0.5)$ diagonal view and the fifth one an equatorial plane view.}
\label{figure:D_3d_snapshots}
\end{figure*}

\begin{figure}[t]
\centering
\includegraphics[width=0.48
\textwidth]{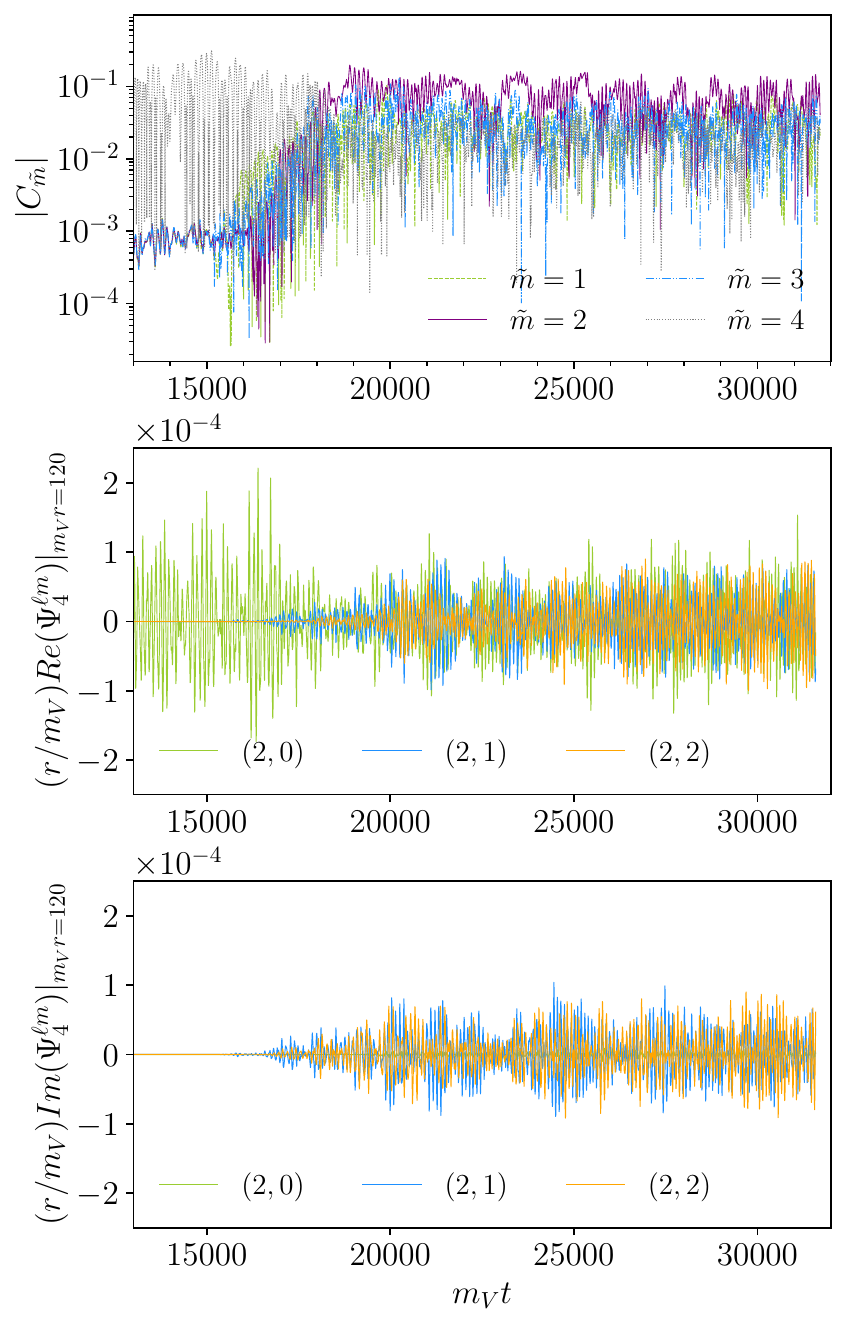}
\caption{As Fig.~\ref{figure:C_azimuthal_modes-gws} but for model $D$.}
\label{D_azimuthal_modes_dominant_gws_modes}
\end{figure}

We have also evolved a model that undergoes a bar-mode instability, exhibiting the same dynamical processes as models $C$ and $E$, but with a completely different outcome. This is our model $D$. For this model the gravitational pull (which drives collapse into a BH in model $C$) and the centrifugal force associated with the bar-mode formation (which leads to fragmentation in model $E$) are finely balanced. As a result, the system settles into a single object with a spheroidal morphology and a pronounced bar-mode structure. This is particularly clear in the last three snapshots of the bottom two rows in Figure~\ref{figure:D_3d_snapshots}. This figure shows snapshots of the energy density of model $D$ (its full evolution is shown in Fig.~\ref{fig:time-evolutions}). We display the initial time of the evolution (first column) and a time before the onset of the bar-mode instability around $m_V t \approx 11800$ (second column). The remaining columns plot the morphology once the bar-mode dominates the evolution. This part of the evolution is clearly visible in the top panel of Fig.~\ref{D_azimuthal_modes_dominant_gws_modes}, with the corresponding gravitational-wave emission plotted in the lower two panels of the same figure. In a similar way to what we observed in model $C$, the spheroidal object in model $D$ experiences a displacement of its center of mass. This shift was significant for rendering the 3D snapshots in the last two rows of Figure~\ref{figure:D_3d_snapshots}, particularly in determining how to clip half of the volume with respect to the new location of the center of mass. 

Correspondingly, the three top rows of Figure~\ref{figure:D_3d_snapshots} display the individual Proca field components (both real and imaginary) with azimuthal numbers $m = 0, -1, -2$, at the same times than for the energy density. From these snapshots we can clearly identify an initial quadrupolar structure corresponding to each component's azimuthal number, noticeable before the onset of the bar-mode instability and prior to an incomplete transition to a multi-$\ell$ Proca star. As in model $E$, this partial transition is caused by the $\tilde{m}=4$ mode, which induces the migration of the $m=|1|$ Proca field component. However, once the bar-mode dominates the evolution and breaks the alignment of the internal rotation axis, the $m=|1|$ and $m=|2|$ constitutive fields end up migrating to an $\ell=1, m=|1|$ configuration. In short, the composite star corresponds to a new configuration only formed by $\ell=1$ Proca fields. 

In order to assess if this new system is still non-rotating, we track the values of the Komar angular momentum given by Eq.~\eqref{eq:Komar_angular_momentum}. We find a zero total angular momentum while the internal rotational axis is still aligned with the $z$ grid axis. Once this is no longer the case, we obtain that the maximum value of angular momentum reaches order $J_{r^*}\sim\mathcal{O}(10^{-2})$, to then decrease by one order of magnitude at the end of the evolution. This result seems to discard that that the system acquires a non-negligible total rotation. However, we cannot claim that the cancellation of the angular momentum of the constitutive fields remains perfectly fine-tuned since the system has suffered odd isometry breakings that can affect the $|m|$ and $-|m|$ Proca fields in different ways.  

\setlength{\tabcolsep}{2.5pt}
\begin{table}[h]
\centering
\begin{tabular}{c c c c c c }
\\ \hline
Model &  \multicolumn{5}{c}{Dynamical process}  \\ \hline\hline
$A$   & ($\ell=2$)\text{-PS} & $\xrightarrow{radial}$ & \text{BH}  \\
$B_1$ & ($\ell=2$)\text{-PS} & $\xrightarrow{\tilde{m}=4}$ & \text{multi-}$\ell$ \text{PS} & $\xrightarrow{radial}$      & \text{BH} \\
$B_2$ & ($\ell=2$)\text{-PS} & $\xrightarrow{\tilde{m}=4}$ & \text{multi-}$\ell$ \text{PS} & $\xrightarrow{radial}$      & \text{BH} \\
$C$   & ($\ell=2$)\text{-PS} & $\xrightarrow{\tilde{m}=4}$ & \text{multi-}$\ell$ \text{PS} & $\xrightarrow{\tilde{m}=2}$ & \text{BH} \\ 
$D$   & ($\ell=2$)\text{-PS} & $\xrightarrow{\tilde{m}=4}$ & (\text{multi-}$\ell$ \text{PS})$^*$ & $\xrightarrow{\tilde{m}=2}$ & ($\ell=1$)-\text{PS}$^*$   \\
$E$   & ($\ell=2$)\text{-PS} & $\xrightarrow{\tilde{m}=4}$ & (\text{multi-}$\ell$ \text{PS})$^*$ & $\xrightarrow{\tilde{m}=2}$ & Binary PS \\ \hline \\ 
\end{tabular}
\caption{Dynamical process for all 2-Proca stars (PS) models studied in this work.}
\label{table:2-Proca_configurations_fate}
\end{table}

Finally, Table~\ref{table:2-Proca_configurations_fate} summarizes all the dynamical processes identified in the long-term evolutions of all the models of our sample. We indicate the transitional stages found and the potential instabilities that trigger them. We also highlight some of the stages with an asterisk to show that in those cases the system cannot be associated with a pure multi-$\ell$ Proca star but undergoes a mode-mixing with a $\tilde{m}=4$ mode. For model $D$ the asterisk also points out that the final state of this multi-field star is not a spherical ($\ell=1$)-Proca star, but an axially symmetric multi-field Proca star with all five constitute fields having a dipolar structure within $\ell=1$.







\section{Conclusion}
\label{sec:conclusions}

In this paper we have tested the non-linear stability properties of $\ell$-Proca stars through long-term numerical-relativity simulations of a sample of spherically symmetric static equilibrium configurations. These solutions incorporate the expected multi-state nature of quantum bosonic fields through their construction as multi-field classical Proca stars with angular momentum in their inner constitutive fields. Such models were built in a previous work~\cite{Lazarte:2024a} and they have provided the initial data for the simulations reported here. Ultimately, this study has been motivated by the astrophysical potential that self-gravitating Proca fields have demonstrated in various scenarios in recent years,  particularly as exotic compact objects that challenge the BH paradigm currently dominating the interpretation of gravitational-wave observations. In this study we have focused on the evolution of ($\ell=2$)-Proca stars. This choice was because evolutions of ($\ell=0$)-Proca stars were already conducted in~\cite{herdeiro2024non} and equilibrium configurations of spherically symmetric ($\ell=1$)-Proca stars were discarded as physically viable bosonic stars because they present a matter singularity at the origin~\cite{Lazarte:2024a}. As in~\cite{herdeiro2024non}, our numerical-relativity simulations were performed in 3-dimensional Cartesian coordinates, without imposing any spatial symmetry. 


Our results have shown that 2-Proca stars are dynamically unstable under generic perturbations in all their domain of existence and show different evolution end-points according to the kind of instability they undergo. As expected, models at the left of the maximum mass configuration in the domain of existence of 2-Proca stars, suffer a radial instability and collapse promptly forming a static Schwarzschild BH. Thus, we have confirmed that such models lie in the radially unstable branch of 2-Proca stars. The rest of our models are  configurations in the expected radially stable branch. Their evolutions have shown that they remain for longer times oscillating  radially and keeping their initial spherical morphology. However, they develop non-axisymmetric instabilities leading to a non-rotating  hexadecapolar energy density distribution, dominated by a $\tilde{m}=4$ mode according to the azimuthal Fourier decomposition of the total energy density of the stars. This instability is triggered by the inevitable small perturbation introduced by the Cartesian grid discretization which, in turn, grows exponentially during a linear regime and saturates when becoming non-linear. 
The new energy density distribution reflects the breaking of the initial spherical symmetry of our stars.

During the non-linear phase of the $\tilde{m}=4$ instability, our models experience a migration to a new multi-field Proca star configuration. This process is a consequence of the change from a quadrupolar to a dipolar angular structure of three ($m=0,|1|$) of the five constitutive Proca fields of the stars. We interpret this as a effective change of their angular momentum numbers from $\ell=2$ to $\ell=1$ while preserving their azimuthal numbers $m$. Since the other two Proca fields with $m=|2|$ maintain their quadrupolar structure defined by $\ell=m=2$, two different angular momentum numbers coexist within the star. The new configurations have been identified as multi-$\ell$ Proca stars. In this new regime, the stars have a spheroidal energy density, dominated by $\ell=1$, with the still-present inner hexadecapolar energy density distribution. Despite this, our more compact configurations exhibit a high degree of axial symmetry, supported by the emission of gravitational waves with only plus polarization. 

Our results also show that these new multi-field configurations are also unstable, leading to two possible outcomes. On the one hand, we have identified a radially unstable branch with configurations suffering a radial collapse into BHs. On the other hand, some models are prone to developing bar-mode instabilities ($\tilde{m}=2$) coming from the constitutive ($m=|2|$)-Proca fields. In turn, different end-points are possible for the latter. We have observed three possible fates listed in ascending order according to the compactness of the stars: {\it (i)} a non-axisymmetric collapse into a BH, {\it (ii)} an additional migration of the ($m=|2|$)-Proca fields from quadrupolar to dipolar configurations, and {\it (iii)} the total fragmentation of the star into a binary Proca star driven by the centrifugal force arising from the rotating bar mode.

Special attention deserves the evolution of model $D$, which begins with a star composed by five Proca fields with quadrupolar structure ($\ell=2$) and ends up as a new solution for which all fields have migrated to the dipolar one ($\ell=1$). This outcome is in agreement with the idea that dipolar Proca stars are dynamically stable and preferred in single-field scenarios, as reported in~\cite{herdeiro2024non,Sanchis_Gual_2019fae,DiGiovanni_2020}. This work, thus, extends such proposal to the multi-field context in the case of five fields, suggesting that it could also be true for three Proca fields. Indeed, our multi-$\ell$ Proca star solutions not contaminated by mode-mixing with the $\tilde{m}=4$ mode showed that the constitutive fields with $m=0,|1|$ started their migration at the same time, which backs up the fact that they seem to behave as a compound system inside the whole star. This raises the possibility that non-spherically symmetric ($\ell=1$)-Proca stars may exist and exhibit strong dynamical stability. Searching for equilibrium configurations and conducting time evolutions of these stars will be the next step in our efforts to find stable multi-field Proca stars that share the stability properties of ($\ell=1$, scalar)-boson stars, allowing us to construct a multi-field family of Proca stars comprising up to three fields where axially symmetric Proca stars are the symmetry-enhanced and unique stable solutions.


\acknowledgments

We thank Eugen Radu for the insightful discussion on the physical relevance of solutions with discrete symmetries, Milton Ruiz for useful comments about numerical arrangements and convergence regimes, and Fabrizio Venturi for helping us with 3D visualization. This work has been supported by the Generalitat Valenciana through a Santiago Grisolía Grant (CIGRIS/2022/164). NSG acknowledges support from the Spanish Ministry of Science, Innovation and Universities via the Ram\'on y Cajal programme (grant RYC2022-037424-I), funded by MCIN/AEI/ 10.13039/501100011033 and by ``ESF Investing in your future”. This work is further supported by the Spanish Agencia Estatal de Investigaci\'on (Grant PID2021-125485NB-C21) funded by MCIN/AEI/10.13039/501100011033 and ERDF A way of making Europe, by the Generalitat Valenciana (Prometeo grant CIPROM/2022/49), 
and by the European Horizon Europe Staff Exchange (SE) programme HORIZON-MSCA2021-SE-01 Grant No.~NewFunFiCO-101086251. This work is also supported by the
Portuguese Foundation for Science and Technology (FCT – Fundação para a Ciência e
a Tecnologia) through the FCT project 2024.05617.CERN (\url{https://doi.org/10.54499/2024.05617.CERN}). The authors acknowledge the computer resources provided by the Red Espa\~nola de Supercomputaci\'on (Tirant, MareNostrum5, and Storage5) and the technical support from the IT departments of the Universitat de Val\`encia and the Barcelona Supercomputing Center (allocation grants RES-FI-2023-1-0023, RES-FI-2023-2-0002, RES-FI-2024-2-0012, and RES-FI-2024-3-0007).




\bibliographystyle{apsrev4-2}
\bibliography{referencias}


\end{document}